# Mechanism of reconstitution/activation of the soluble PQQ-dependent glucose dehydrogenase from *Acinetobacter calcoaceticus*: a comprehensive study

Claire Stines-Chaumeil,[*,a] Francois Mavré,[b] Brice Kauffmann,[c] Nicolas Mano,[a] and Benoit Limoges[*,b]

[a] CNRS, Université de Bordeaux, CRPP, UMR5031, 115 Avenue Schweitzer, F-33600 Pessac, France.

[b] Université de Paris, Laboratoire d'Electrochimie Moléculaire, UMR 7591, CNRS, F-75013 Paris, France.

[c] CNRS UMS3033, INSERM US001, Université de Bordeaux, IECB, 2, Rue Robert Escarpit, F-33607 Pessac, France.

## Abstract

The ability to switch on the activity of an enzyme through its spontaneous reconstitution has proven to be a valuable tool in fundamental studies of enzyme structure/reactivity relationships or in the design of artificial signal transduction systems in bioelectronics, synthetic biology, or bioanalytical applications. In particular, those based on the spontaneous reconstitution/activation of the apo-PQQ-dependent soluble glucose dehydrogenase (sGDH) from *Acinetobacter calcoaceticus* were widely developed. However, the reconstitution mechanism of sGDH with its two cofactors, *i.e.* the pyrroloquinoline quinone (PQQ) and $Ca^{2+}$, remains unknown. The objective here is to elucidate this mechanism by stopped-flow kinetics under single-turnover conditions. The reconstitution of sGDH exhibited biphasic kinetics, characteristic of a square reaction scheme associated to two activation pathways. From a complete kinetic analysis, we were able to fully predict the reconstitution dynamic, but also to demonstrate that when PQQ first binds to the apo-sGDH, it strongly impedes the access of $Ca^{2+}$ to its enclosed position at the bottom of the enzyme binding site, thereby greatly slowing down the reconstitution rate of sGDH. This slow calcium insertion may purposely be accelerated by providing more flexibility to the $Ca^{2+}$ binding loop through the specific mutation of the calcium coordinating P248 proline residue, reducing thus the kinetic barrier to calcium ion insertion. The dynamic nature of the reconstitution process is also supported by the observation of a clear loop shift and a reorganization of the hydrogen bonding network and van der Waals interactions observed in both active sites of the apo and holo forms, a structural change modulation that was revealed from the refined X-ray structure of apo-sGDH (PDB:5MIN).

## Highlights

- A comprehensive reconstitution mechanism of the PQQ-dependent sGDH is proposed
- When the PQQ cofactor first binds to the apo-sGDH, it strongly impedes the access of $Ca^{2+}$
- Mutation of the P248 residue allows for an increasing accessibility of $Ca^{2+}$ to the PQQ-occupied site

1. Introduction

Many enzymes require the noncovalent insertion of cofactors (*e.g.*, flavins, hemes, metal ions, iron-sulfur clusters) into their apoprotein binding site to be fully active. In numerous cases, this enzyme activation process, also termed enzyme reconstitution, can be achieved *in vitro* through a precise control of the reaction conditions (temperature, buffer composition, pH …), paving thus the way for fundamental investigations of enzyme structure/reactivity relationships (notably by taking advantage of engineered binding partners such as apoenzyme mutants and/or cofactor analogs)[1,2,3,4,5,6,7,8] or even for discovering new enzyme functionalities.[9,10,11] Besides to the capacity of switching on the activity of an enzyme by simply adding its cofactor, enzyme reconstitution has also been advantageously exploited in different biotechnological applications,[5,10,12,13,14,15,16,17,18,19] ranging from the design of artificial signal transduction systems for analytical purposes to the development of new applications in biotechnology, bioelectronics or synthetic biology.[10,20,21,22,23,24,25]

Among the enzymes whose catalytic activity can be easily and efficiently switched on *via* reconstitution, the soluble quinoprotein glucose dehydrogenase (sGDH, code UniprotKB F0KFV3) is certainly the most prevailing and attractive.[15,26,27,28,29,30,31,32,33] The reason for such interest is that, in the presence of calcium ions, the catalytic property of sGDH for aldoses oxidation can be rapidly and spontaneously activated through the specific and tight binding of its PQQ cofactor to the apoprotein (apo-sGDH). This property has led to the design of novel analytical methods for the sensitive detection of calcium ions[30] or PQQ,[29] and to the conception of unique signal amplification strategies to boost the analytical performances of miscellaneous affinity binding assays.[15,31,32] Other features which makes sGDH an attractive activatable enzyme is the ease with which the apoenzyme can be overproduced in a recombinant strain of *Escherichia coli* and isolated with a high yield and purity (totally free of PQQ).[34,35] The reconstitution reaction is furthermore spontaneous and fast in the presence of $Ca^{2+}$, mainly driven by the high-affinity binding of PQQ to the apo-sGDH (equilibrium dissociation constant in the sub-nanomolar range).[29] Also, the holoenzyme exhibits a remarkably high catalytic activity towards the oxidation of glucose with concomitant reduction of a wide range of natural or artificial electron acceptors, a catalytic reactivity that can be easily monitored either spectrophotometrically or electrochemically. Albeit there are several works reporting on sGDH reconstitution[34,36,37,38] as well as on its exploitation in different biotechnological applications,[15,29,30,31,32] its reconstitution mechanism remains unknown. Previous studies have yet established that, similarly to alcohol dehydrogenase, reconstitution of the homodimeric sGDH enzyme requires the binding of six calcium ions and two PQQ molecules.[39,40,41] Four of the $Ca^{2+}$ ions are involved in the functional dimerization of the two protein subunits (*i.e.*, held together by four $Ca^{2+}$ ions shared at the interfaces) and two for the activation of each PQQ cofactor present in each of two apoenzyme subunits (*i.e.*, one $Ca^{2+}$ ion per PQQ in close interaction within the binding site).[36] Other divalent cations ($Cd^{2+}$, $Mn^{2+}$, $Sr^{2+}$) were also revealed effective in the dimerization and activation of the apoenzyme.[37] In addition, the binding of PQQ was shown remarkably efficient not only thermodynamically but also kinetically. Dissociation constants in the nano- to picomolar range were reported in the presence of millimolar $Ca^{2+}$, demonstrating the high-affinity binding of PQQ to the apo-sGDH,[29,37] while a second order rate constant of ~1-2 × $10^6$ $M^{-1}·s^{-1}$ characterizing the binding of PQQ to the apo-sGDH was determined either for the enzyme reconstituted in homogenous solution[42] or once immobilized on an electrode surface.[29] These kinetics were however examined in the presence of a large excess of $Ca^{2+}$ and under steady-state kinetics, preventing thus a detailed understanding of the role played by this second cofactor in the reconstitution mechanism.

The aim of the present paper is thus to fill this gap by investigating the reconstitution mechanism of apo-sGDH with PQQ and $Ca^{2+}$ by stopped-flow kinetics. From analysis of the transient kinetics under single-turnover conditions, we show that the reconstitution follows two different pathways, each depending on the order the cofactors are positioned in the binding site, *i.e.* PQQ binding followed by $Ca^{2+}$ or $Ca^{2+}$ binding followed by PQQ. It finally allows us to propose a comprehensive overview of the reconstitution mechanism of sGDH, from which we were then able to quantitatively rationalize and predict the reconstitution dynamics. Moreover, from determination of the complete X-ray structure of apo-sGDH and comparison to the holo-sGDH, we were able to evidence a movement of the loop located near the binding pocket, highlighting thus the dynamic nature of the reconstitution process.

## 2. Material and Methods

*2.1 Reagents*

PQQ, glucose, TRIS buffer, $CaCl_2$, dichlorophenolindophenol (DCPIP), phenazine methosulfate (PMS) and other chemicals were purchased from SIGMA and used as received.

*2.2 Expression, purification and reconstitution of wild-type and mutant sGDHs from A. calcoaceticus*

A QuickChange site-directed mutagenesis kit (Stratagene) was used for the mutations. The presence of mutations was verified by DNA sequencing at the functional genomic center of Bordeaux. Wild-type apo-sGDH and P248A apo-sGDH were produced, purified and then reconstituted with PQQ and $Ca^{2+}$ as reported earlier.[43] Apo-sGDH and $holo_{ox}$ protein concentrations were assessed spectrophotometrically (Varian Cary 100) from the absorbance at 277 nm (for the extinction coefficients see Table S1 in Supporting Information).[39] $Holo_{red}$ is obtained after addition of glucose. Enzymes activities were measured using standard protocols.[43] In this manuscript, the concentration of enzymes is unless otherwise stated expressed per monomer (or subunit). After purification and reconstitution apo and holo forms of the enzyme were both stored at -80°C in a 50 mM TRIS buffer (pH 7.5) containing 3 mM $CaCl_2$.

*2.3 UV-visible stopped flow kinetics under single-turnover*

Pre-steady-state kinetic analysis was carried out on a MOS 450 stopped-flow apparatus (Biologic) and the data analyzed either using the Biokine software package or the Origin software. The dead-time of the stopped-flow apparatus was determined to be 3 ms using standard protocols.[44] All the experiments were performed in triplicate with three different batches of enzymes. For each experiment, an average of at least 4 runs was performed.

*2.4 Influence of $CaCl_2$ pre-incubated in a syringe containing apo-sGDH or PQQ*

To check if there was any effect of calcium three sets of experiments were performed: (i) 10 μM PQQ in syringe 1 – 5 μM apo-sGDH subunit + 6 mM $CaCl_2$ in syringe 2; (ii) 10 μM PQQ +

6 mM CaCl$_2$ in syringe 1 – 5 µM apo-sGDH subunit in syringe 2; (iii) 10 µM PQQ + 3 mM CaCl$_2$ in syringe 1 – 5 µM apo-sGDH + 3 mM CaCl$_2$ in syringe 2. Control experiments showed that the enzyme remains dimeric upon dilution.

*2.5 Kinetic studies of the tryptophan fluorescence quenching*

Wavelengths of excitation and emission were determined at the equilibrium by performing fluorescence quenching with a Xe lamp in the presence of 5 µM apo-sGDH subunit in presence or absence of 10 µM PQQ in a 50 mM TRIS buffer (pH 7.5) at 10°C. The time-course of sGDH fluorescence quenching induced by the binding of PQQ was recorded on a thermostated MOS 450 Biologic stopped-flow apparatus equipped with fluorescence detector and a XeHg lamp (at 800 V). Absorption and emission wavelengths were set at 297 nm and 340 nm, respectively. At such wavelengths, there was no fluorescence contribution from either the buffer or PQQ. For the kinetic experiments, the first syringe was filled with 5 µM of apo-sGDH subunits and the second one with PQQ and CaCl$_2$ at different concentrations ranging from 5 to 40 µM for PQQ and from 9 µM to 150 mM for CaCl$_2$.

*2.6 Formation rate of holo$_{red}$ from holo$_{ox}$*

The reduction rate of holo$_{ox}$ by glucose was monitored at 338 nm with the stopped-flow apparatus operating in absorbance mode (XeHg lamp at 200 V) thermostated at a controlled temperature of 10°C. One syringe was filled with 5 µM of holo$_{ox}$ subunit in a 50 mM TRIS buffer + 3 mM CaCl$_2$ (pH 7.5), and the other one with 200 µM of glucose in a 50 mM TRIS buffer + 3 mM CaCl$_2$ (pH 7.5). The final concentrations of sGDH and glucose after mixing are divided by 2.

*2.7 Formation rate of holo$_{red}$ from wild-type or P248A apo-sGDH*

The rate of PQQ reduction by glucose was measured at 338 nm with the stopped-flow apparatus operating in absorbance mode (XeHg lamp at 200 V) thermostated at controlled temperature of 10°C. One syringe was filled with 5 µM apo sGDH subunits and the other one with a mixture of glucose, PQQ and CaCl$_2$ at concentrations ranging from 10 to 80 µM for PQQ and from 18 µM to 300 mM for $CaCl_2$, the concentration of glucose being fixed at 200 µM. The final concentrations of sGDH, PQQ, glucose and $CaCl_2$ after mixing are divided by 2.

*2.8 Structure determination and refinement*

*Crystallization.* Single crystals of the apo-sGDH from *Acinetobacter calcoaceticus* have been grown by the vapor diffusion method. These crystals diffract to beyond 1.76 Å and are suitable for X-ray crystallography on a home source rotating anode. The space group was determined to be triclinic P(1). One asymmetric unit contains a dimer of the apo-sGDH molecule. Preliminary screening of crystallization conditions was carried out using the Jena Bioscience Classic Kit (number 2). Crystals suitable for crystallography experiments were directly obtained by the sitting-drop method (in 96 well-plate) with drops consisting of 0.5 µL protein solution (at 15 mg/mL) and 0.5 µL reservoir solution equilibrated against 70 mL well solution at 293 K, using a reservoir solution composed of 30% (w/v) PEG 4000, 200 mM $CaCl_2$ and 100 mM HEPES (pH 7.5) (condition D2).

*Data Collection.* X-ray diffraction data of apo-sGDH were collected at the IECB X-Ray facility (CNRS UMS3033, INSERM US001, University of Bordeaux) at 100 K with a 3 kW Rigaku FR-X X-ray generator equipped with a Hybrid Dectris Pilatus 200K detector. XDS and XSCALE[45] were used for data integration and scaling. Statistics of data collection are listed in Table S2.

*Structure Determination and Refinement.* The crystal structure of apo-sGDH was determined to 1.76 Å resolution with phases determined by molecular replacement (1C9U as starting model) using PHASER.[46] Model building was performed with Coot and illustrated with PyMOL.[47] Crystallographic refinement was performed with programs REFMAC[48] and BUSTER-TNT.[49] Coordinates have been deposited with PDB accession code 5MIN.

*2.9 Kinetic simulations*

The open-source software COPASI 4.11 has been used for numerical kinetic simulations of the reconstitution mechanism shown in Scheme 2.

## 3. Results

As previously reported by Duine and coll.,[34,36] the isolated apo-sGDH from *E. coli* recombinant strain is in the dimeric form, containing thus $Ca^{2+}$ at the interface between the two subunits but no PQQ. The preparation also remains dimeric after desalting over gel filtration or upon high dilution in a $Ca^{2+}$-free buffer. Even in the presence of an excess of EDTA the protein remains as a dimer, demonstrating that $Ca^{2+}$ once bounded at the subunit interface is locked.

The kinetics of sGDH reconstitution was investigated by stopped-flow experiments. The methodology consists to rapidly mix PQQ, apo-sGDH, $Ca^{2+}$, and excess glucose, and then to follow by UV-visible spectroscopy the transient reduction of PQQ once the latter is properly bounded and activated into the apoenzyme binding site. By this way, single-turnover reduction of PQQ witnesses the active behavior of the reconstituted enzyme since free PQQ cannot be directly chemically reduced by glucose in solution. The reduction rate of the PQQ cofactor in the enzyme was monitored at $\lambda$ = 338 nm, a wavelength which relates to a maximal absorbance change between the oxidized (*i.e.*, holo$_{ox}$) and reduced state of sGDH (*i.e.*, holo$_{red}$ wherein the PQQ cofactor is doubly reduced in its pyrroloquinoline quinol form or PQQH$_2$) (Figure S1 in Supporting Information). In order to properly extract the kinetics information on the enzymatic reconstitution process, it was indispensable to select experimental conditions that lead to an almost instantaneous reduction of the bounded (and activated) PQQ cofactor relative to the rate of its upstream equilibrium binding to the apoenzyme binding site (in other words, to select conditions where the reduction of the holo$_{ox}$ by glucose is not rate-limiting). To evaluate how fast is the reduction rate of the wild type holo$_{ox}$ by an excess glucose, 2.5 μM subunits of native sGDH were mixed with 100 μM glucose in a stopped-flow apparatus and the transient kinetic monitored by UV-visible spectroscopy at 338 nm (Figure S2). The fast absorbance change that occurs at 338 nm over a maximal variation of $\Delta A_{max}$ ~ 0.05 (corresponding to a full reduction of sGDH) and within a time scale shorter than 30 ms allows setting the time window required for complete enzyme reduction by glucose (it is worth to note that because of the fast rate of sGDH reduction a large fraction of the absorbance change was completed during the dead time of the instrument). From the known maximal absorbance change at 338 nm and the fit of an exponential law to the experimental data, an apparent first order rate constant of 130 ± 2 s$^{-1}$ was found. This value agrees well to that previously determined for the reductive half-reaction of sGDH[42,39] under similar conditions, confirming that the catalyzed hydride transfer reaction from glucose to the PQQ cofactor in the active site is a fast process. These experimental conditions (*i.e.*, 2.5 µM subunits of sGDH and 100 µM glucose), finally leading to a fast reduction of holo$_{ox}$ by glucose, were then selected for the further kinetic experiments.

To investigate the kinetics of sGDH reconstitution under single-turnover conditions, PQQ concentrations ranging from substoichiometric to overstoichiometric ratios (*i.e.*, from 0.5 to 40 µM) relative to the total amount of available apo-sGDH binding sites (2.5 µM subunits) were tested in the presence of fixed concentrations of CaCl$_2$ (3 mM) and glucose (100 µM). The data reported in Figure 1A exhibit biphasic exponential kinetics within a time frame extending from 20 ms to 100 s, much slower than the time scale required for the above reduction of the holoenzyme by glucose (half-life $t_{1/2}$ ~ 5 ms, Figure S2). It therefore means that under the selected conditions, the reaction is essentially rate-controlled by the equilibrium binding of PQQ to the apoenzyme and not by the following step of holoenzyme reduction by glucose. This result also shows that the proposed transient UV-visible kinetic methodology is relevant for characterizing the reconstitution mechanism of sGDH.

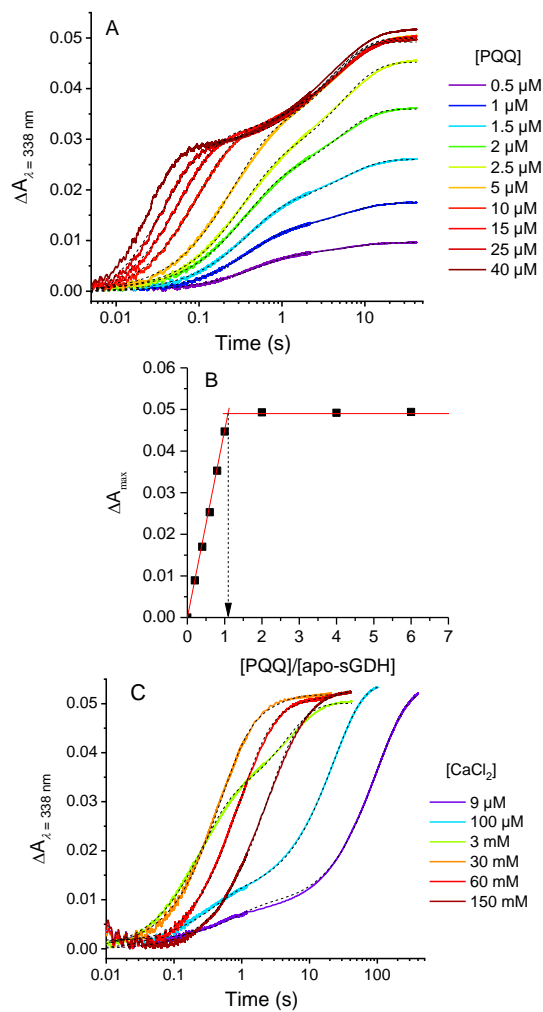

**Figure 1.** Stopped-flow transient traces of the catalytic reduction of PQQ by glucose during the reconstitution/activation of apo-sGDH into holo-sGDH. (A) Kinetics traces (average of 4 consecutive experiments) obtained after mixing 2.5 μM subunits of apo-sGDH with 100 μM glucose, 3 mM $CaCl_2$ and different concentrations of PQQ (from left to right): 40, 25, 15, 10, 5, 2.5, 2, 1.5, 1, and 0.5 μM (the code color for each [PQQ] is reported on the graph). All concentrations given are after mixing. The reactions were monitored at 338 nm and 10°C in a 50 mM TRIS (pH 7.5). (B) Titration plot of apo-GDH by PQQ obtained from the plot of total absorbance change in A as a function of the [PQQ]/[apo-sGDH] ratio. (C) Kinetics traces (average of 4 consecutive experiments) obtained after mixing 2.5 μM subunits of apo-sGDH with 5 μM PQQ (2-fold excess), 100 μM glucose and different concentrations of $CaCl_2$: 9 μM, 100 μM, 3 mM, 30 mM, 60 mM, and 150 mM (the code color for each [$CaCl_2$] is reported on the graph). All concentrations given are after mixing and other experimental conditions are the same than in A. The black dotted curves in A and C are the best fits of a bi-exponential law to the experimental data.

The maximal amplitude of absorbance change at 338 nm ($\Delta A_{max}$) in Figure 1A is linked to the fraction of apo-sGDH that reconstitutes and activates. By reporting this maximal amplitude as a function of the [PQQ]/[apo-sGDH] ratio, the titration plot of apo-sGDH by PQQ can be obtained (Figure 1B). The plot shows that the apo-GDH can be fully reconstituted (and so fully activated) at a 1.1:1 ratio of PQQ/apoenzyme subunits, which is, within the experimental error, close to that expected for the binding of one PQQ molecule per subunit. Consequently, the two well distinct kinetics phases observed in Figure 1A suggests the existence of two different enzyme activation pathways, occurring in parallel but at different pace.

The kinetics of sGDH reconstitution was also examined for different concentrations of calcium (ranging from 9 µM to 150 mM), keeping constant the ratio of PQQ/apo-sGDH subunits to an excess of 2 (Figure 1C). The gradual transition from a biphasic kinetics at low calcium concentrations (for which the slow kinetics pathway predominates) to a simple first-order kinetics at high $Ca^{2+}$ concentration provides evidence that calcium plays a key role in the partitioning of reactants between the two parallel pathways. In addition, the almost identical values of maximal absorbance change recorded at the end of kinetics ($\Delta A_{max} \sim 0.05$) are in good agreement with the expected burst amplitude for a quantitative activation of the 2.5 µM subunits of apo-sGDH present in solution.

Given that both PQQ and $Ca^{2+}$ are required for enzyme activation, the reconstitution mechanism can thus be reasonably described by a square reaction scheme with two parallel pathways in competition, each depending on the binding order of the two entities to form an activated ternary complex (*i.e.*, $holo_{ox}$), which then rapidly reduces to $holo_{red}$ in the presence of glucose (Scheme 1). From the amplitude of absorbance change of one phase ($\Delta A_1$ or $\Delta A_2$) relative to the total absorbance change ($\Delta A_{max} = \Delta A_1 + \Delta A_2$), these experiments give an access to the relative population of the different transient species contributing to each reaction path. The amplitudes $\Delta A_1$ and $\Delta A_2$ as well as the observed rates $k_{1,obs}$ and $k_{2,obs}$ associated to each kinetic phase were recovered from the nonlinear regression fit of a biexponential function to the kinetic traces as shown in Figures 1A and 1C (see Tables S1 and S2 for the extracted data). Under the reaction conditions of 3 mM $Ca^{2+}$ (Figure 1A), the ratio $\Delta A_1/\Delta A_{max}$ is almost the same regardless of PQQ concentration and close to an average value of 0.6 (Table S3), a result which thus provides evidence for a roughly constant balanced initial distribution of sGDH between two populations that can each independently reconstitute according to a fast or a slow pathway. This partitioning between fast and slow reconstitution is also strongly dependent on $[Ca^{2+}]$ as attested by the plot of $\Delta A_1$ or $\Delta A_2$ as a function of calcium concentration in Figure S3 (the graph was obtained from the values reported in Table S4 as well as those collected from other experiments), showing an hyperbolic dependence of the relative amplitude of each phase on $[Ca^{2+}]$. This behavior provides evidence for a fast equilibrated binding reaction between apo-sGDH and $Ca^{2+}$ (leading to formation of apo-sGDH/Ca$^{2+}$), fast enough to be assumed always at equilibrium before PQQ significantly binds to the protein. From the fit of a standard binding isotherm to the data (Figure S3), an apparent equilibrium dissociation constant of $K_d^{\text{apo-sGDH/Ca}} = 2.0 \pm 0.50$ mM was recovered.

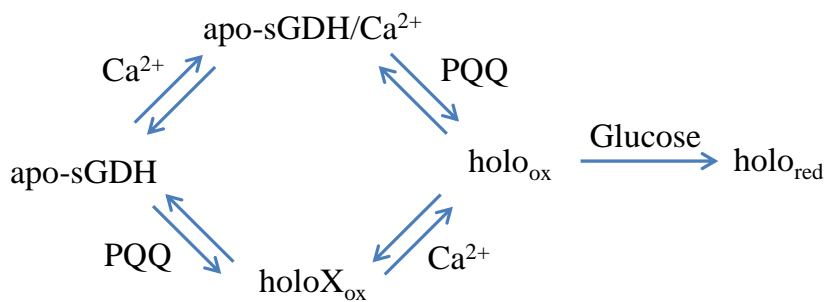

**Scheme 1.** Square scheme reconstitution mechanism of apo-sGDH with the random binding of Ca$^{2+}$ and PQQ to form the ternary activated complex holo$_{ox}$, which then rapidly converts into holo$_{red}$ in the presence of glucose.

The values $k_{1,\text{obs}}$ and $k_{2,\text{obs}}$ gathered in Tables S1 and S2 are plotted in Figure 2 as a function of PQQ concentration. In conditions of pseudo-first order (ratio [PQQ]/apo-sGDH ≥ 2) the observed rate $k_{1,\text{obs}}$ shows a linear dependency with [PQQ], suggesting that the fast phase of enzyme reconstitution is rate-limited by the entrance of PQQ. In contrast, $k_{2,\text{obs}}$ is found independent of [PQQ] (average value of 0.2 s$^{-1}$) over the whole explored range (0.5 to 40 µM), demonstrating that PQQ is not the rate-limiting factor in this slow enzyme reconstitution pathway. Interestingly, the observed rates were also found independent of the order the three partners were mixed inside the two-syringe stopped-flow device (Table S6), showing similar values of $k_{1,\text{obs}}$ and $k_{2,\text{obs}}$ whether the content of the syringe loaded with apo-sGDH or PQQ was pre-incubated or not with Ca$^{2+}$. This observation reinforces the hypothesis that a fast equilibrium binding occurs between apo-sGDH and Ca$^{2+}$, fast enough to be nearly always at equilibrium during the whole enzyme kinetic experiments.

The linear relationship between $k_{1,\text{obs}}$ and [PQQ] at high PQQ/apo-sGDH ratios (*i.e.*, conditions of pseudo-first-order in apoenzyme) gives an access to the second order rate constant associated to the binding of PQQ to the apo-sGDH under 3 mM Ca$^{2+}$. From the linear regression fit in Figure 2 (red line) a $k_{1,\text{PQQ}}$ value of $(1.08 \pm 0.05) \times 10^6$ M$^{-1}$ s$^{-1}$ is obtained. This value was confirmed from the analysis of a larger set of kinetics experiments performed under pseudo-first order conditions (see Figure S4), leading to $k_{1,PQQ} = (1.12 \pm 0.01) \times 10^6$ M$^{-1}$ s$^{-1}$.

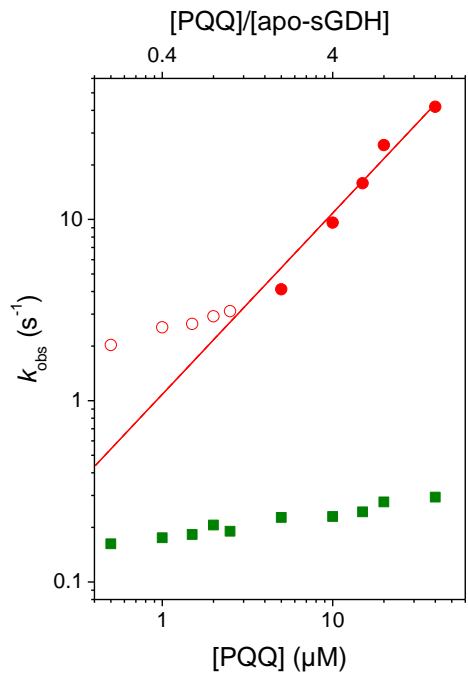

**Figure 2.** Plots of (red dots) $k_{1,obs}$ and (green squares) $k_{2,obs}$ (both recovered from the fits of a biexponential function to experimental kinetics curves shown in Figure 1A and whose data are compiled in Table S3) as a function of PQQ concentration or [PQQ]/[apo-sGDH] ratio. The straight line is the linear regression fit to the $k_{1,obs}$ data determined under pseudo-first order conditions in apoenzyme (*i.e.*, [PQQ] $\geq 5$ µM). The slope of the linear regression fit allows to calculate the bimolecular rate constant of the reaction between PQQ and apo-sGDH, leading to a value of $k_{1,PQQ} = (1.08 \pm 0.05) \times 10^6$ M$^{-1}$ s$^{-1}$.

It was previously shown by Duine and coll[37] that the addition of PQQ to apo-sGDH can be monitored from the quenching of fluorescence engendered by the six tryptophan residues present on the protein. We therefore took advantage of this quenching of fluorescence to study the kinetics of sGDH reconstitution. Figure S5 shows that the maximum of fluorescence emitted by apo-sGDH at 340 nm (red curve) is strongly quenched after the addition of PQQ together with Ca$^{2+}$ ions (blue curve). Stopped flow experiments were thus performed at 340 nm by monitoring the quenching rate of the apo-sGDH fluorescence as a function of [PQQ] and [Ca$^{2+}$] (only overstoichiometric ratios of [PQQ]/[apo-sGDH] were considered because of a too weak and noisy fluorescence under substoichiometric conditions) (Figure 3). In contrast to the above kinetics traces determined from the absorbance change at 338 nm, the time course plots of quenching of fluorescence in Figure 3A and 3B are all well-characterized by a single kinetic phase, whatever the concentration of PQQ and $Ca^{2+}$. From the fit of a monoexponential decay to the highest PQQ concentrations, the recovered observed rate not only linearly depends on the PQQ concentration (Figure S4, red triangles), but also coincide with the values of $k_{1,obs}$ previously determined from the absorbance change (it is to note that the significant change observed for both the fluorescence offset and the overall fluorescence amplitude as a function of PQQ concentration is presumed associated to a fast non-specific quenching of fluorescence generated by excess PQQ in solution). This result strongly supports that the quenching of fluorescence of apo-sGDH in the absence of glucose addresses the same fast kinetic process than the one associated to the fast phase of absorbance change at 338 nm in the presence of glucose, a fast process we have attributed to the entrance of PQQ in the apo-sGDH binding site. From the slope of the linear regression fit to the data in Figure S4, a second order rate constant of $(1.25 \pm 0.03) \times 10^6$ $M^{-1} \cdot s^{-1}$ was obtained, which is thus close to the one obtained from the absorbance change at 338 nm (*i.e.*, $1.12 \pm 0.01 \times 10^6$ $M^{-1} \cdot s^{-1}$).

Concerning the effect of $[Ca^{2+}]$ on the time course of fluorescence quenching (performed here at a fixed [PQQ]/[apo-sGDH] ratio of 2), the kinetic plots in Figure 3B show that $k_{obs}$ remains relatively constant and independent of $[Ca^{2+}]$ ($k_{obs} \sim$ 4-5 $s^{-1}$) until the latter remains lower than 3 mM, while it progressively decreases for $Ca^{2+}$ concentrations > 3 mM (Table S5). This outcome correlates very well with the fast kinetic phase of absorbance change at 338 nm, wherein $k_{1,obs}$ is similarly slowed down at $[Ca^{2+}]$ > 3 mM (the kinetic traces performed at 30, 60 and 150 mM $Ca^{2+}$ in Figure 3B matched quite well those reported in Figure 1C at the same $Ca^{2+}$ concentrations). As it will be discussed latter on, we have attributed this behavior to an inhibition of the fast enzyme reconstitution pathway by excess $Ca^{2+}$. Additionally, the fact that the kinetics of fluorescence quenching follows a single exponential decay independent of $[Ca^{2+}]$ for concentrations lower than 3 mM, provides evidence that the surveyed process under these conditions is the conformational changes induced by the binding of PQQ to the apoenzyme site (and this, whether or not there is a $Ca^{2+}$ ion present in the binding site). Such a behavior clearly differs from the dependence of $k_{1,obs}$ on $[Ca^{2+}]$ observed in Figure 1C and Table S4, which likewise addresses the binding of PQQ to the apo-sGDH but differently since in this case this is the fraction of apo-sGDH (*i.e.*, apo-sGDH/$Ca^{2+}$) that is involved into the fast activation pathway (a fraction that itself depends on the calcium concentration through the fast equilibrated reaction between apo-sGDH and $Ca^{2+}$, leading thus to a dependence of $k_{1,obs}$ on $[Ca^{2+}]$ at low $Ca^{2+}$ concentrations).

Duine and his collaborators have demonstrated that, in the absence of calcium ion, the apo-GDH can reconstitute into a much less active form of sGDH, baptized holoX$_{ox}$, a $Ca^{2+}$-free enzyme able to slowly convert into its reduced holoX$_{red}$ counterpart in the presence of glucose.[34,36] In order to characterize the kinetics of these two processes (*i.e.* the rate of holoX$_{ox}$ formation as well as the rate of the slow holoX$_{ox}$ reduction by glucose in the absence of calcium), stopped flow kinetics experiments were performed without calcium and in the presence of a large excess of EDTA (in order to scavenge any residual traces of free $Ca^{2+}$ in solution) (Figure S9). The conversion rate of apo-sGDH into holoX$_{ox}$ was extracted from the quenching of fluorescence of tryptophan at 340 nm as a function of different concentrations of PQQ. The resulting kinetic traces in Figure S9 are characterized by an exponential decay of fluorescence, similar to that previously observed in the presence of calcium. The observed rate extracted from each plot varies linearly with excess PQQ (Figure S9), thus allowing to recover a second order rate constant of $k_{2,PQQ} = 0.94 \times 10^6$ $M^{-1} \cdot s^{-1}$. This value is very close to the rate $k_{1,PQQ}$ determined in the presence of calcium, leading us to conclude that the binding rate of PQQ to the apoenzyme is almost unchanged whether there is or not a calcium ion anchored within the binding site of apo-sGDH.

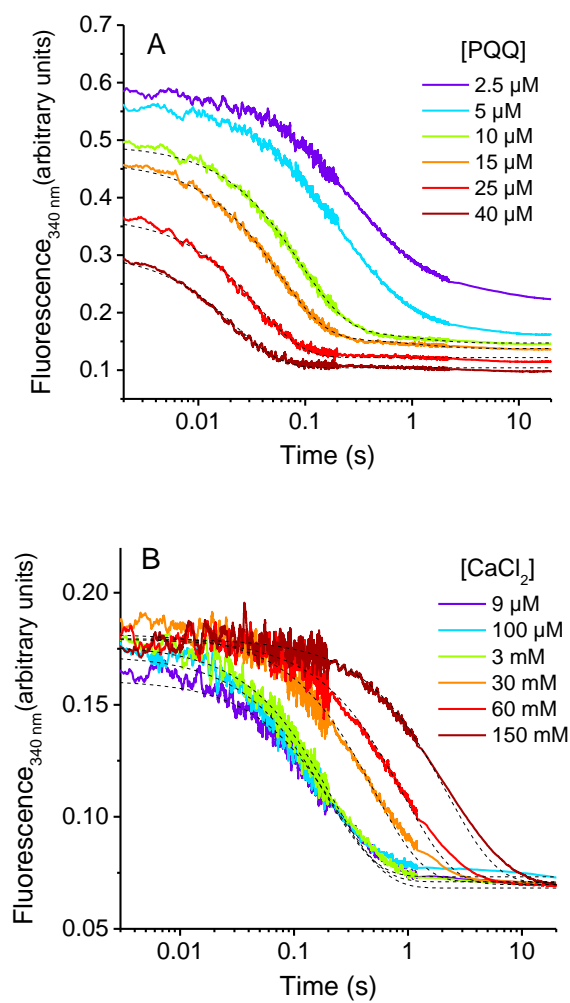

**Figure 3.** Transient traces of the quenching of tryptophan fluorescence residues of apo-sGDH during enzyme reconstitution with PQQ and calcium. (A) Stopped-flow kinetics obtained after mixing 2.5 µM subunits of apo-sGDH with 3 mM $CaCl_2$ and different concentrations of PQQ (from left to right): 40, 25, 15, 10, 5 and 2.5 µM. All concentrations given are after mixing. The reactions were carried out at 10°C in a 50 mM TRIS (pH 7.5) and the fluorescence monitored at 340 nm with an excitation at 297 nm. (B) Stopped-flow kinetics obtained after mixing 2.5 µM subunits of apo-sGDH with 5 µM PQQ and different concentrations of $CaCl_2$ (from left to right): 9 µM, 100 µM, 3 mM, 30 mM, 60 mM, and 150 mM. All concentrations given are after mixing and other experimental conditions are the same than in A. The black dotted curves in A and B are the best fits of a mono-exponential function to the experimental data.

To further characterize the reduction rate of $holoX_{ox}$ into $holoX_{red}$ subsequent to the addition of glucose, the reaction was monitored by UV-visible spectrophotometry at 307 nm (see Figure S6 for the selection of this wavelength) in a stopped-flow apparatus. The kinetic trace reported in Figure S7 demonstrates a very slow reduction rate of $holoX_{ox}$ by glucose, leading to a pseudo-first order rate of $k_{red,X} = 8.4 \times 10^{-4}$ s$^{-1}$, a value that is more than 160 000 times slower than in the presence of calcium ($k_{red} = 130$ s$^{-1}$, Figure S2). This result confirms the strong effect of $Ca^{2+}$ on the sGDH reactivity, the role of which is to activate the PQQ cofactor (activation at the $C_5$ position of PQQ) and to promote the hydride transfer reaction between glucose and PQQ.[50]

The possibility to isolate the holoX$_{red}$ form of sGDH has pushed us to determine the incorporation rate of $Ca^{2+}$ into holoX$_{red}$ to form the holo$_{red}$ – a binding reaction wherein the reduced PQQ cofactor already present in the enzyme pocket is assumed to hinder the access of calcium, the latter position of which is located just behind the PQQ binding site at the bottom of the active site.[40] For such a purpose, we have monitored by stopped-flow the transition from holoX$_{red}$ to holo$_{red}$ by simply mixing different concentrations of $Ca^{2+}$ to a holoX$_{red}$ solution followed by the absorbance measurement at 338 nm. The reaction rate was observed to hyperbolically depends on the concentration of $Ca^{2+}$ (Figure S8), suggesting a two-step binding reaction mechanism that we have assumed to obey a Michaelis-Menten-type kinetic. From the best curve fitting, the following maximal rate ($k'_{3,Ca}$) and apparent Michaelis-Menten constant ($K^M_{3,Ca}$) were recovered: $k'_{3,Ca} = 0.52 \pm 0.01$ s$^{-1}$ and $K^M_{3,Ca} = 5.30 \pm 0.69$ mM.

On account of the very slow reduction of holoX$_{ox}$ into holoX$_{red}$ and the fast and almost irreversible conversion of apo-sGDH to holoX$_{ox}$ (the strong equilibrium binding constant is assumed to lead in a slow dissociation rate of PQQ from holoX$_{ox}$, *vide infra*), it makes it possible to characterize the binding reaction of $Ca^{2+}$ to holoX$_{ox}$ (in a similar way as above with the holoX$_{red}$), generating thus the transient species holo$_{ox}$ which then rapidly converts into holo$_{red}$ in the presence of glucose. To assess the rate of this reaction, different concentrations of calcium in the presence of excess of glucose were then mixed with the holoX$_{ox}$ in a stopped-flow apparatus and the reaction mixture then kinetically monitored by UV-vis at 338 nm (Figure S10). The resulting kinetic plots were fitted to a monoexponential function leading to an observed rate that hyperbolically depends on the concentration of $Ca^{2+}$ (Figure S10). Again this behavior suggests a two-step binding reaction mechanism that we have once more assumed obeying to a Michaelis-Menten-type reaction. From the best curve fitting, the following parameters were obtained: $k'_{2,Ca} = 1.19 \pm 0.2$ s$^{-1}$ and $K^M_{2,Ca} = 0.6 \pm 0.3$ mM. The value of $k'_{2,Ca}$ (that can formally be assimilated to $k_{2,obs}$) is here somewhat higher when compared with $k_{2,obs}$ determined in Figure 2. We have attributed this discrepancy to the high scattering we have encountered in the experimental determination of $k'_{2,Ca}$ and $K^M_{2,Ca}$ (Figure S10), probably resulting from the presence of a large excess of EDTA. In addition, one may note that the value of $K^M_{2,Ca}$ is an order of magnitude lower than $K^M_{3,Ca}$, revealing the strongest interaction of calcium with the oxidized form of PQQ as compared to the reduced one.

To provide further evidence that the slow activation pathways in the sGDH reconstitution mechanism is linked to the difficulty for calcium to reach its enclosed position when PQQ is already present in the enzyme binding site (steric effect), we have mutated the P248 amino-acid into alanine. The mutation of this proline was chosen because it is located within a loop near the active site that is expected to significantly contribute in the 3D organization of the enzyme active site (the distinctive cyclic structure of proline's side chain is well-known to provide a conformational rigidity compared to other amino acids, affecting thus the secondary structure of proteins near a proline residue[51]). An additional reason is because the two carbonyl groups present on the amino-acids P248 and G247 are known to directly interact with the calcium ion, which itself are in strong interaction with PQQ (Figure 4D). The P248A mutation is thus anticipated to significantly increase the flexibility and mobility of the loop coordinating the calcium ion, therefore to potentially reduce the kinetic cost of the calcium insertion when PQQ binds first. The reconstitution of the mutated P248A apo-sGDH was then characterized by stopped-flow kinetic experiments under the same conditions than for the wild-type apoenzyme (Table S6). Likewise to the wild-type enzyme, biphasic kinetic traces were obtained leading for the fast phase to an observed rate constant of $k_{1,obs} = 12 \pm 2$ s$^{-1}$ (Table S6), which is within a range of a factor 2 comparable to that obtained with the wild-type apo-sGDH (*i.e.*, $k_{1,obs} = 5.4 \pm 0.9$ s$^{-1}$). This result demonstrates a negligible effect of the mutation on the binding rate of PQQ. In contrast, the rate of the slow phase is observed significantly enhanced for the mutant enzyme ($k_{2,obs} = 3.7 \pm 0.8$ s$^{-1}$, which is ~16-fold higher than for the wild-type, *i.e.* $k_{2,obs} = 0.23 \pm 0.02$ s$^{-1}$) (Table S6), suggesting that the P248A mutation has a substantial structural effect on the accessibility of Ca$^{2+}$ to its enclosed position within the holoX active site. As anticipated, the mutation leads to an increased flexibility of the loop coordinating Ca$^{2+}$, which thereby facilitates the incorporation and/or access of the calcium ion within the sterically congested site of holoX.

Finally, with the aim to fully characterize the environment of the binding pocket before reconstitution, we have determined the X-ray structure of apo-sGDH in the presence of calcium at 1.76 Å of resolution (PDB 5MIN). In the previously published X-ray structure of apo-sGDH, two loops from position 105 to 110 and 333 to 346 and a disulfide bridge (CYS345-CYS338) were lacking (PDB 1QBI),[41] while in our new X-ray structure of apo-sGDH the disulfide bridge is intact and the two loops are clearly defined (see Figure 4A and B for the loop from position 334 to 346 including the disulfide bridge and Figure S11 to see the electron density of the loop 333 to 346). Interestingly, the superimposition of the apo and holo Cα traces shows a shift of 1 to 2 Å in the relative position of the loop located from 333 to 346 (Figure 4B). It clearly suggests that a local conformational rearrangement occurs during and/or following the binding of PQQ, which might act as a lock to ensure PQQ's strong affinity (dissociation constant of 30 pM and dissociation rate as low as $k_{-1,PQQ} = 3.4 \times 10^{-5}$ s$^{-1}$, *vide infra*). It might then results in a kinetic barrier (or a steric constraint) to overcome for the insertion of Ca$^{2+}$ when PQQ is already in place in the apo-sGDH binding site.

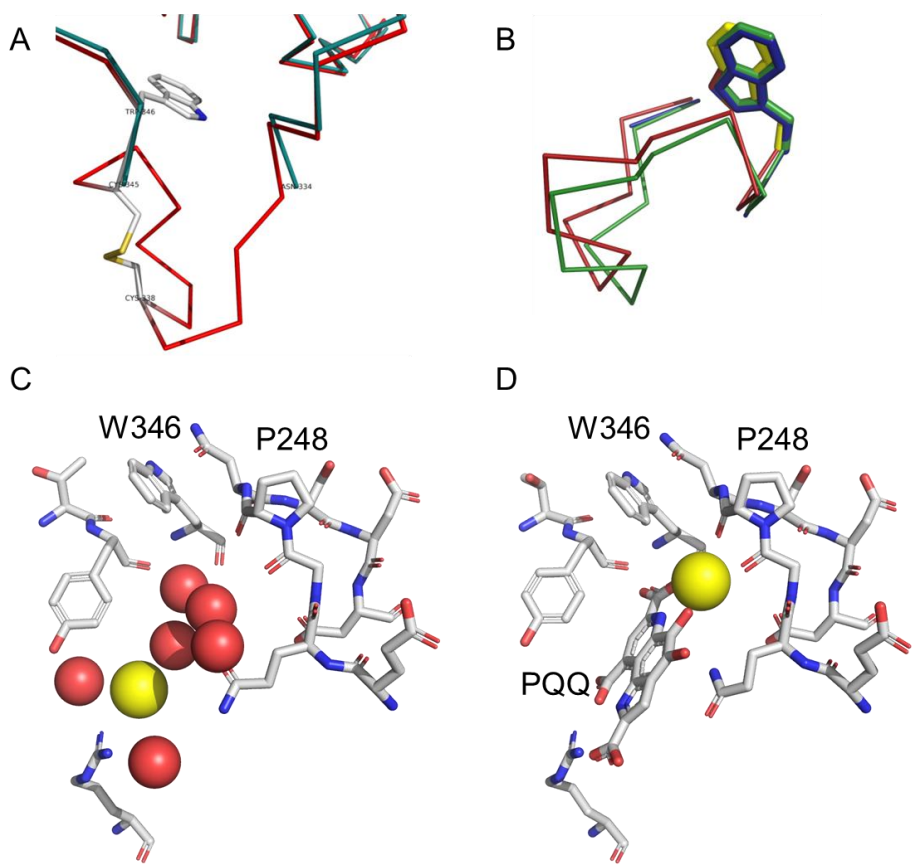

**Figure 4.** (A) Superimposition of the X-ray structures of (red) apo-sGDH (PDB:5MIN) and (green) holo-sGDH (PDB:1QBI) in the loop region ASN334 to TRP346, showing the disulfide bridge (CYS345-CYS338) and the trace of the main chain (missing in 1QBI). (B) Superimposition of the Cα traces of apo-sGDH (5MIN, red), holo-sGDH (1C9U, green) and apo-sGDH (1QBI, blue) structures in the same region showing the loop shifting (overall RMSD 1.13 Å, 1.9 Å for GLY339). The TRP346 (in stick mode) position is well conserved. Same view of the key residues and cofactors in the active site of apo-sGDH (PDB:5MIN) (C) and holo-sGDH (PDB:1C9U) showing the different position of the calcium atoms in both structures and the water molecule network filling the cavity in the absence of PQQ in the apo form.

## 4. Discussion

On the basis of the overall results obtained above, we can finally propose a more elaborate version of the minimum square reaction mechanism described in Scheme 1 by the mechanism depicted in Scheme 2. The latter involves a preferred-order binding of the two cofactors rather than a completely random mechanism (a preferred-order binding that strongly depends on the calcium concentration) to form, on one side (fast pathway), a ternary holo$_{ox}$ complex that rapidly converts into its reduced holo$_{red}$ form in the presence of glucose, and, on the other side (slow pathway), a binary holoX$_{ox}$ complex that can either (*i*) be slowly converted in the presence of $Ca^{2+}$ into a ternary one according to a two-step reaction (assumed here analogous to a Michaelis-Menten-type process wherein $Ca^{2+}$ reacts first through a fast equilibrium binding followed by a slow first order reaction to lock $Ca^{2+}$ in its sterically congested site in the enzyme) followed by fast reduction with glucose, or (*ii*) be directly very slowly reduced by glucose into holoX$_{red}$ before to definitely converts into holo$_{red}$ through a two-step reaction with $Ca^{2+}$ (that we have again considered as a Michaelis-Menten type process in agreement with the hyperbolic rate dependence with [$Ca^{2+}$]). In addition, to take into account the reversible apparent inhibition observed at high $Ca^{2+}$concentrations (> 3 mM) (*i.e.*, decrease of $k_{1,obs}$ at high [$Ca^{2+}$] – see Table S5), an additional equilibrium reaction was added (inhibition path) by considering the fast reversible formation of a PQQ/$Ca^{2+}$ complex, which thus has the consequence to decrease the amount of free PQQ in solution and so to slow-down (or inhibit) the enzymatic reconstitution at very high [$Ca^{2+}$]. Although there is no information about the real nature of this inhibition process, we have postulated that it may arise from an equilibrium binding interaction between PQQ and $Ca^{2+}$ (with kinetic rate constants $k_i$ and $k_{-i}$), leading thus to a PQQ/$Ca^{2+}$ complex ineffective to reconstitute the apo-sGDH into a holo active form. The interaction between PQQ and $Ca^{2+}$ is supported by previous works where it has been shown that a PQQ/$Ca^{2+}$ complex can be generated at high $Ca^{2+}$ concentration in either an organic solvent[52] or an aqueous buffer.[53]

Full kinetic description of the global reconstitution mechanism in Scheme 2 implies the definition of 14 rate constants, which have to be as far as possible independently determined. Let's first focus on the fast pathway. Five rate constants are required for this pathway: two for the reversible binding of $Ca^{2+}$ into the apo-sGDH (*i.e.*, $k_{1,Ca}$ and $k_{-1,Ca}$ for the association and dissociation rate constants of $Ca^{2+}$, respectively), two for the binding of PQQ to the apo-sGDH/$Ca^{2+}$ complex (*i.e.*, $k_{1,PQQ}$ and $k_{-1,PQQ}$ for the association and dissociation rate constants of PQQ, respectively), and one for the reduction of holo$_{ox}$ to holo$_{red}$ at a saturated concentration of glucose (*i.e.*, $k_{red}$ = 130 s$^{-1}$). As discussed earlier, the binding of $Ca^{2+}$ is considered very fast since preincubation of calcium with either PQQ and/or apo-sGDH does not influence or change the experimental kinetic traces. Considering the relative small hydrodynamic size of the calcium ion and the broad accessibility of the apo-sGDH binding pocket (the binding site of PQQ is located at the top of the barrel in a deep, broad and positively charged cleft[40]), we can legitimately assume that the second order binding rate of $Ca^{2+}$ to apo-sGDH is diffusion-controlled, which therefore means a rate constant $k_{1,Ca}$ in the range of $10^8$-$10^9$ $M^{-1} \cdot s^{-1}$.[54] Kinetic of the reverse reaction can be set from the knowledge of the equilibrium dissociation constant. The latter can be estimated from the change in absorbance of the fast pathway relative to the slow pathway as a function of $[Ca^{2+}]$ (Figure S3), a ratio which set the fast equilibrium between the two pathways (slow and fast). From the analysis of this ratio, an equilibrium constant in the millimolar range was found (*i.e.*, $K_d^{apo-sGDH/Ca}$ = 5.3 mM). It follows that $k_{-1,Ca}$ is in the range of $10^5$-$10^6$ $s^{-1}$. This agrees with the fact $Ca^{2+}$ is only coordinated by the two carbonyl oxygens of P248 and G247. The fast binding of PQQ to the apo-sGDH/$Ca^{2+}$ was determined from analysis of the observed rate of the fast kinetic phase as a function of [PQQ]. A bimolecular rate constant of $k_{1,PQQ}$ = 1.1 × $10^6$ $M^{-1} \cdot s^{-1}$ was obtained from the linear regression fit of $k_{1,obs}$ as a function of [PQQ] (Figure 2), which is a rather fast rate, not far from a diffusion-controlled reaction. The dissociation rate of PQQ from $holo_{ox}$ was then indirectly recovered from the equilibrium dissociation constant between PQQ and apo-sGDH/$Ca^{2+}$ ($K_d^{holo_{ox}}$) we had previously determined at room temperature in the presence of 3 mM $CaCl_2$ (*i.e.*, conditions for which most of the reconstitution goes through the fast pathway).[29] On account of the remarkably low value of $K_d^{holo_{ox}}$ = (3.0 ± 1.5) × $10^{-11}$ M previously published,[29] a rather slow dissociation rate of $k_{-1,PQQ}$ = 3.4 × $10^{-5}$ $s^{-1}$ can be calculated.

Let's now consider the slow pathway for which the initial binding of PQQ to the apo-sGDH leads to the formation of $holoX_{ox}$, followed by the binding of $Ca^{2+}$ to generate the $holo_{ox}$ form. The existence of $holoX_{ox}$ was confirmed from the UV-vis absorption spectra in Figure S6 showing a maximum at 340 nm, which is 12 nm lower than for the $holo_{ox}$ but 12 nm higher than of free PQQ.[36] This was also corroborated by the fast quenching of fluorescence after rapid mixing of the apo-sGDH with PQQ in the absence of $Ca^{2+}$ (Figure S9), signing the fast incorporation of PQQ into the apo-sGDH binding site. Such a behavior is consistent with previous works showing that $Ca^{2+}$ ions in the binding site is not require for a strong binding of PQQ to the apoenzyme binding site.[50] The calcium ion remains however indispensable for the enzyme activation, because in its absence the $holoX_{ox}$ can be only very slowly reduced into $holoX_{red}$ with glucose as attested by our kinetic experiment in Figure S7, leading to a $k_{red,X}$ value of 8 × $10^{-4}$ $s^{-1}$ (which is ~162 500-fold slower than $k_{red}$) (see Scheme 2). The slow reconstitution pathway of sGDH is expected to predominate at low calcium concentration. For instance, most of the reconstitution goes through the slow pathway at the lowest $Ca^{2+}$ concentration we have tested (*i.e.*, 9 µM). At such a low calcium concentration, the fluorescence quenching is mainly due to the binding of PQQ into the apo-sGDH. In the absence of calcium, the monoexponential decay of fluorescence as a function of [PQQ] (Figure S9) allows to extract the bimolecular rate constant, $k_{2,PQQ}$, characterizing the binding of PQQ into the $Ca^{2+}$-free apo-sGDH site. A value of $(0.94 \pm 0.20) \times 10^6$ $M^{-1} \cdot s^{-1}$ was obtained which is very close to the value of $k_{1,PQQ}$ (*i.e.*, $1.1 \times 10^6$ $M^{-1} \cdot s^{-1}$), demonstrating that the binding rate of PQQ is not significantly affected either there is or not a $Ca^{2+}$ ion present in the enzyme binding site. It however does not inform us on the value of the dissociation rate $k_{-2,PQQ}$, which is expected to be somewhat a little higher than $k_{-1,PQQ}$ because there is no $Ca^{2+}$ in the $holoX_{ox}$ to further stabilize the PQQ cofactor. However, we can reasonably hypothesize that the absence of $Ca^{2+}$ should not decrease the affinity binding of PQQ by a factor greater than one or two orders of magnitude. So, for that reason, we have assumed a $k_{-2,PQQ}$ value lower than $10^{-2}$ $s^{-1}$ (meaning a $K_d^{holoX_{ox}}$ value $> 10^{-8}$ M).

The last rate constants that are needed to fully describe the slow pathway are those associated to the two-step binding of $Ca^{2+}$ to $holoX_{ox}$, characterizing thus the entrance of $Ca^{2+}$ into the $holoX_{ox}$ enzyme site followed by its slow binding at the bottom of the enzyme active site sterically constrained by the presence of the PQQ cofactor. A rough estimation of the two parameters (*i.e.*, $K_{2,Ca}^M$ and $k'_{2,Ca}$) characterizing this two-step Michaelis-Menten-type reaction was obtained from the stopped-flow kinetic experiments reported in Figure S10. The value of $k'_{2,Ca}$ (1.2 $s^{-1}$) is within the range of $k_{2,obs}$ (0.2 $s^{-1}$) we had previously determined from kinetic analysis of the slow enzyme reconstitution pathway in Figure 2. According to Scheme 2, these two rate constants can be assimilated because addressing the same rate-limiting step in the mechanism. It is worth to note that owing to the high scattering in the data of Figure S10, the values of $K_{2,Ca}^M$ and $k'_{2,Ca}$ are entailed by a high degree of incertitude that does not guaranty their high reliability, an issue that may explain the observed divergence between $k'_{2,Ca}$ and $k_{2,obs}$. We have thus finally considered the value of $k_{2,obs}$ inferred from Figure 2 as the more relevant data for characterizing $k'_{2,Ca}$.

Once a comprehensive overview of the overall kinetic rate constants characterizing the global reaction mechanism proposed in Scheme 2 was obtained, it was interesting to see whether this mechanism, with its associated rate constants, could be used to numerically simulate and predict the experimental kinetic plots of Figure 2. For such a purpose we have used the open-source software COPASI which allows solving numerically kinetic models with varying levels of complexity.[55] Using the overall rate constants reported in black and green in Scheme 2 and Table 1 and adjusting the rate constants in blue, we were finally able to predict quite well by simulation the UV-visible experimental kinetic traces given in Figure 2 (see Figure 5 for a side-by-side comparison of simulated curves to the experimental ones). This good agreement between simulations and experiments was made possible thanks to the iterative adjustment of $K_{2,Ca}^{M}$, $k_{2,Ca}^{'}$, and $K_{d}^{PQQ/Ca}$ to values of 0.1 mM, 0.25 s$^{-1}$, and 15 mM, respectively, and also by taking into consideration the individual absorbance of each of the different intermediate species generated along the reaction (this has been achieved by using the extinction coefficients gathered in Table S1 and considering the numerically simulated time-course concentration profiles of the overall species adsorbing at 338 nm). The rather excellent agreement between the simulated kinetic and experimental curves *a posteriori* strongly supports the proposed mechanism.

Finally, the slow pathway seems to be the consequence of the presence of PQQ in the enzyme binding site that then hampers, through a steric effect, the access of Ca$^{2+}$ to its enclosed position. This difficulty to enter the site while PQQ is already bound to the protein can be better apprehended in Figure 6, wherein Ca$^{2+}$ lies at the bottom of the active site, while PQQ is located above. This steric effect is experimentally supported by the result we have obtained with the P248A mutant (Table S6), showing that the rate-limiting step of the slow enzyme reconstitution pathway (characterized by $k_{2,Ca}^{'}$) can be significantly enhanced by reducing the rigidity (or increasing the flexibility) of the enzyme binding site. In addition, comparison of the X-ray structures of apo- and holo-sGDH evidences the dynamic nature of the reconstitution process since it suggests that a local rearrangement of loop 333-346 (close to the binding pocket) can occur upon PQQ binding and correct positioning of Ca$^{2+}$ to switch on the activity (Figure 4). It is therefore possible that modulation of the overall dynamic process through steric constraints would be at the origin of the observed kinetic differences between the two pathways.

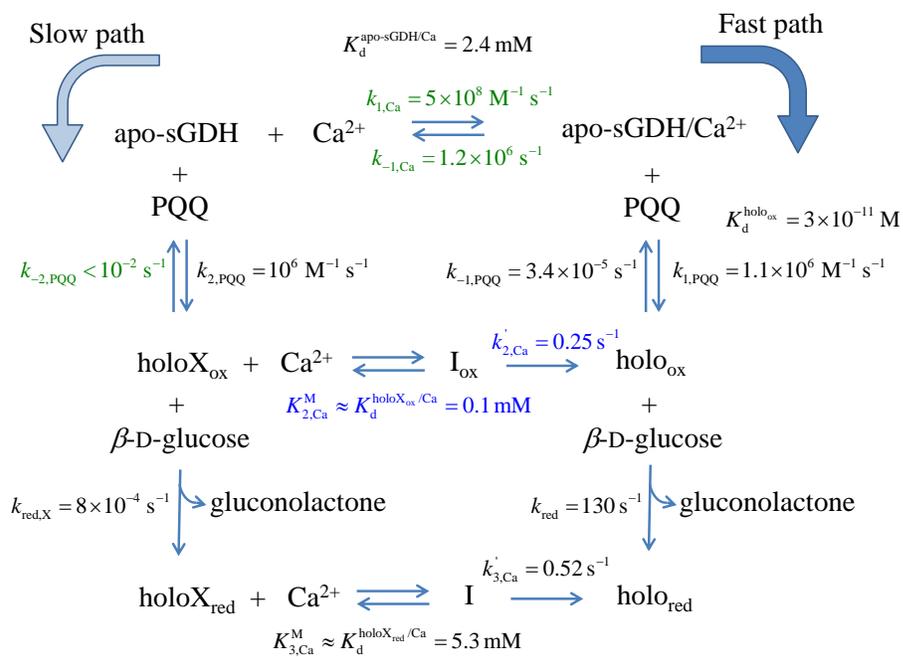

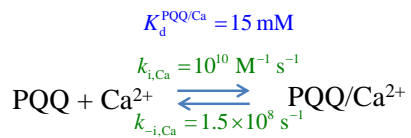

**Scheme 2.** Global mechanism for the reconstitution/activation of apo-sGDH into active PQQ-sGDH.

**Table 1.** Rate constants determined (in black) directly from the experiments or (in blue) indirectly from the best fitting between the numerical kinetic simulations and the experimental kinetic plots. The others rate constants (in green) were estimated on the assumption that $Ca^{2+}$ can bind to the enzyme according to a diffusion-controlled reaction.

| Reaction | Equilibrium constants | Rate constants [c] |
|---|---|---|
| apo-sGDH + $Ca^{2+}$ ⇌ apo-sGDH/$Ca^{2+}$ | $K_d^{apo\text{-}sGDH/Ca}$ = 2.0 mM | $k_{1,Ca}$ = 5 × 10$^8$ M$^{-1}$ s$^{-1}$; $k_{-1,Ca}$ = 1.2 × 10$^6$ M$^{-1}$s$^{-1}$ |
| apo-sGDH/$Ca^{2+}$ + PQQ ⇌ holo$_{ox}$ | $K_d^{holo_{ox}}$ = 3.0 × 10$^{-11}$ M | $k_{1,PQQ}$ = 1.1 × 10$^6$ M$^{-1}$ s$^{-1}$; $k_{-1,PQQ}$ = 4 × 10$^{-5}$ s$^{-1}$ |
| apo-sGDH + PQQ ⇌ holoX$_{ox}$ | | $k_{2,PQQ}$ = 0.94 × 10$^6$ M$^{-1}$ s$^{-1}$ |
| holoX$_{ox}$ + $Ca^{2+}$ ⇌ I$_{ox}$ ⟶ holo$_{ox}$ [a] | $K_{2,Ca}^M$ = 0.6 mM [a] | $k'_{2,Ca}$ = 1.2 s$^{-1}$ [a] |
| | $K_{2,Ca}^M$ = 0.1 mM [a, b] | $k'_{2,Ca}$ = 0.25 s$^{-1}$ [a, b] |
| holoX$_{red}$ + $Ca^{2+}$ ⇌ I ⟶ holo$_{red}$ [a] | $K_{3,Ca}^M$ = 5.3 mM [a] | $k'_{3,Ca}$ = 0.5 s$^{-1}$ [a] |
| holo$_{ox}$ + glucose ⟶ holo$_{red}$ | | $k_{red}$ = 130 s$^{-1}$ |
| holoX$_{ox}$ + glucose ⟶ holoX$_{red}$ | | $k_{red,X}$ = 8 × 10$^{-4}$ s$^{-1}$ |
| PQQ + $Ca^{2+}$ ⇌ PQQ/$Ca^{2+}$ | $K_d^{PQQ/Ca}$ = 15 mM | $k_{i,Ca}$ = 10$^{10}$ M$^{-1}$ s$^{-1}$; $k_{-i,Ca}$ = 1.5 × 10$^8$ M$^{-1}$s$^{-1}$ |

[a] Michaelis-Menten-type reaction are characterized by two parameters: the Michaelis constant ($K_i^M$) and the turnover rate ($k'$)

[b] Determined from the best fit of the simulated kinetic traces to the experimental ones (see Figure 5 and text for details)

[c] The absolute rate constants in green color were estimated by assuming a diffusion-controlled bimolecular reaction.

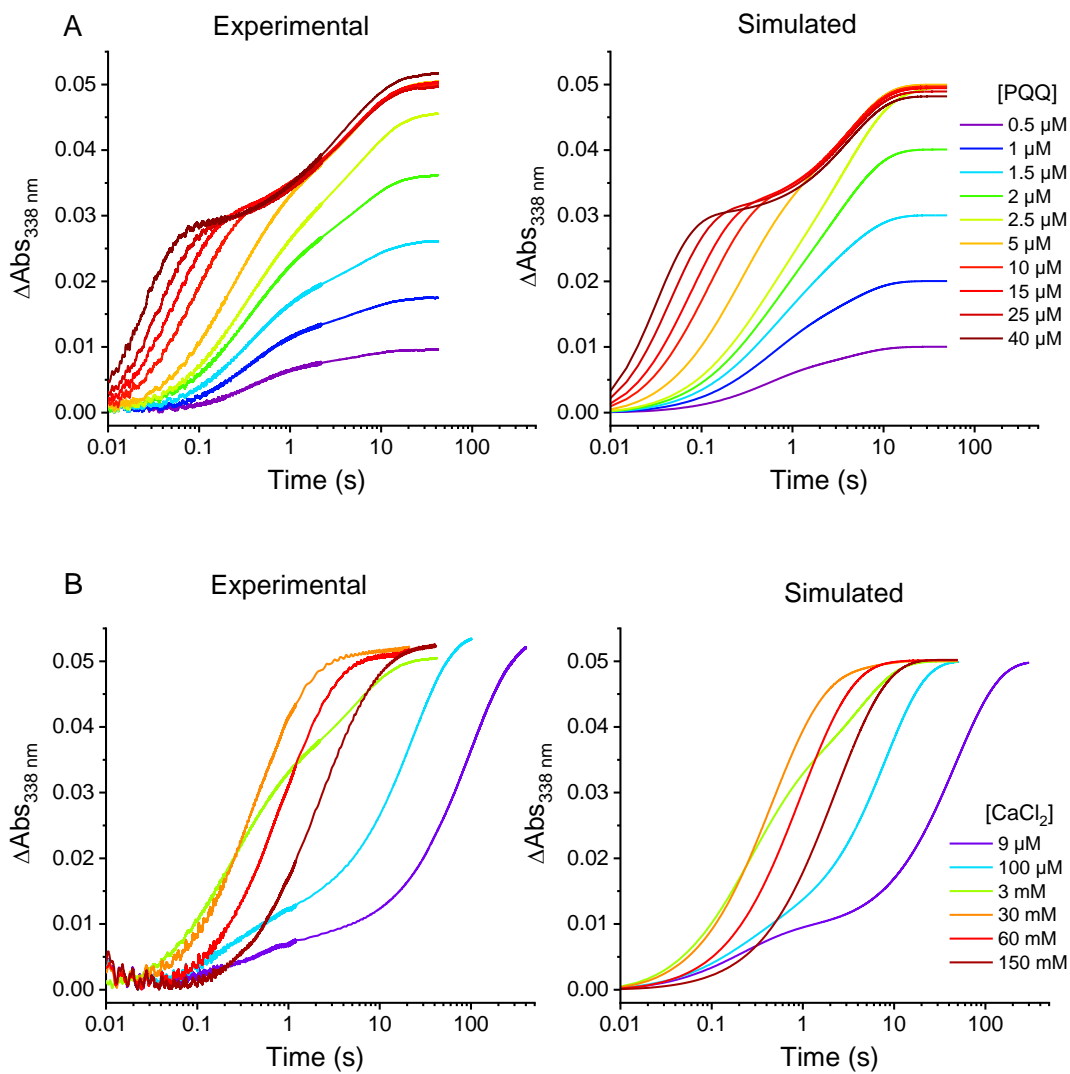

**Figure 5.** Comparison between experimental and simulated UV-vis kinetic traces (monitored at 338 nm). The experimental curves are the same than in Figure 1A and 1C. The simulated curves were obtained from the numerical simulation of the global reaction mechanism shown in Scheme 2 and using the reported rate constants (some of these constants are those determined experimentally, while others were iteratively adjusted in such a way to get the best fit between the simulations and the experimental plots).

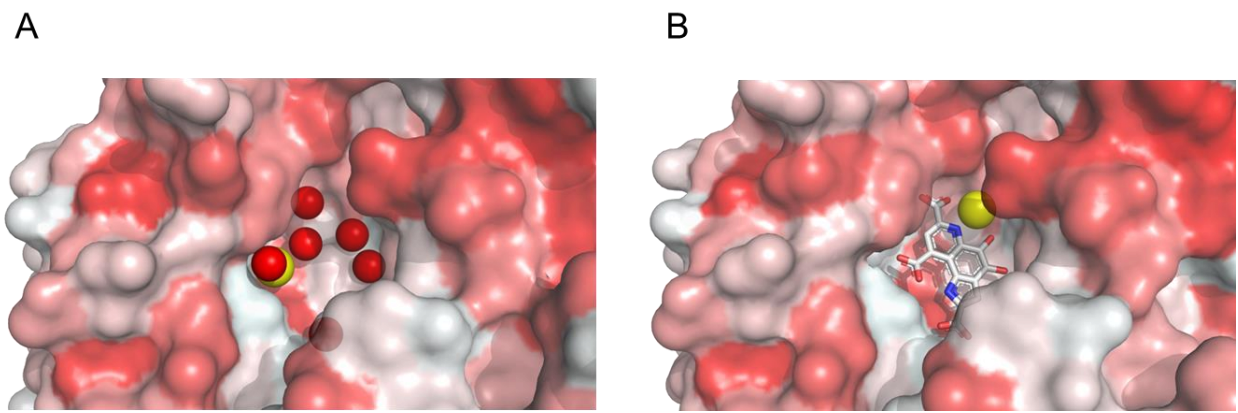

**Figure 6.** Representative views of the opened and solvent accessible cavities in (A) apo-sGDH (5MIN) and (B) holo-sGDH (1C9U) showing the different positions of the Ca atoms, the network of water molecules in the apo-sGDH and illustrating the difficulty that calcium may have in accessing its binding site when PQQ binds first to the protein. The surfaces are colored according to the hydrophobicity of the residues (the more hydrophobic residues in red).

## 5. Conclusion

We presented here a detailed description of the reconstitution mechanism of sGDH from *A. calcoaceticus.* This enzyme provides an ideal playground for the comprehension of the processes though which an enzyme recovers its full activity. It is particularly relevant for enzymes that admit two different partners for a complete activation (here PQQ and $Ca^{2+}$). We have notably shown in this case that the order in which each of the partners binds to the protein strongly impacts the reconstitution kinetics. This might help in the design of biotechnological applications in which the dynamic of such an activity switch is crucial for the response of the system. In addition, thanks to the detailed knowledge of both structure and reactivity, we have also demonstrated that this dynamic may be modulated through directed mutation of the binding site.

**Acknowlegments**

The authors thank la Région Nouvelle Aquitaine for financial support.

**Appendix A. Supplementary data**. Figures S1 to S11 and Tables S1 to S6.

**A**

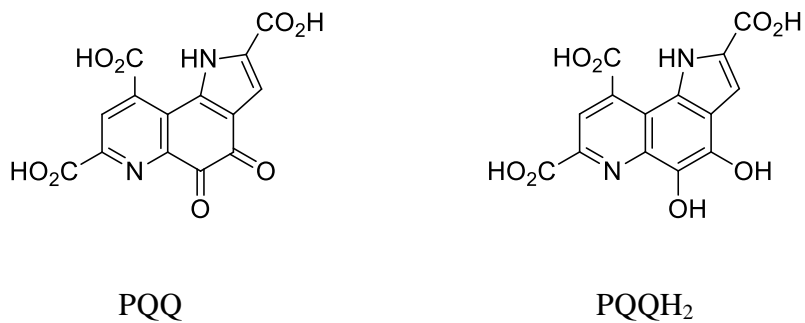

PQQ                  PQQH$_2$

**B**

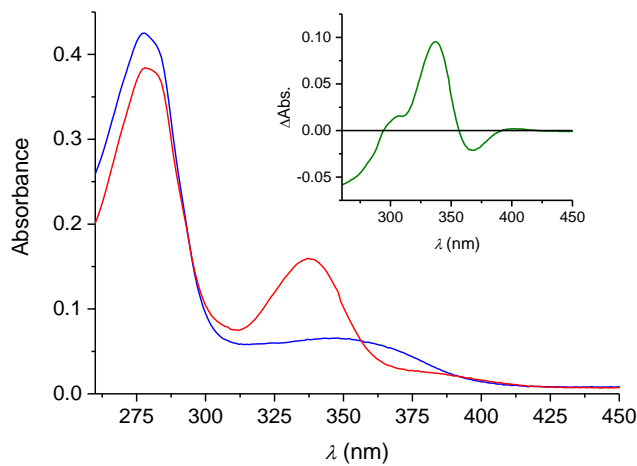

**Figure S1.** A) Chemical structure of PQQ and of its doubly reduced form. B) Absorbance spectra of (blue line) holo$_{ox}$-sGDH (5.0 µM monomeric binding site or subunit) determined after 30 min preincubation at room temperature of apo-sGDH with 2-fold excess of PQQ followed by gel filtration on a PD10 column, and of (red line) holo$_{red}$-sGDH (5.0 µM subunit) after addition of 1 mM D-glucose. The inset show the difference spectrum obtained by substraction of holo$_{ox}$ spectrum from holo$_{red}$ spectrum. The difference of absorbance at 338 nm leads to a difference extinction coefficient of $\Delta\varepsilon_{338}$ = 19 500 M$^{-1}$.cm$^{-1}$.

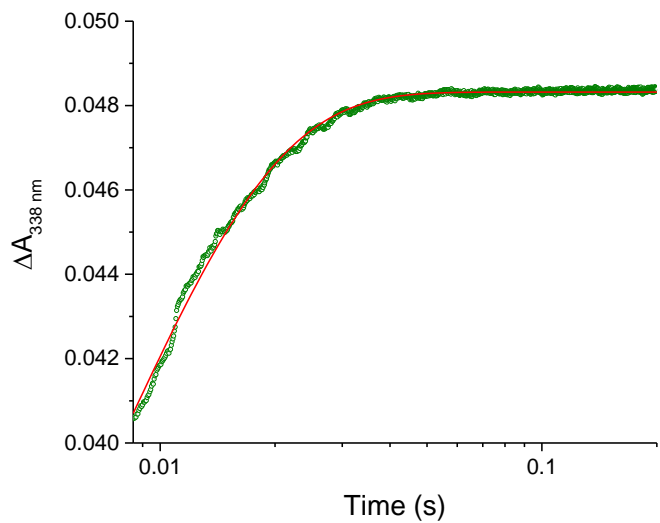

**Figure S2.** Transient kinetics of the reductive half-reaction of the native sGDH by D-glucose. The sGDH (wild-type $holo_{ox}$ form, final concentration 2.5 µM subunit) and D-glucose (final concentration 100 µM), both prepared in a 50 mM TRIS buffer containing 3 mM $CaCl_2$ (pH 7.5), were mixed in the stopped flow apparatus ($T$ = 10°C) and the absorbance monitored at 338 nm (corresponding to the conversion of $holo_{ox}$ into $holo_{red}$). The smooth red curve represents the fit of the data to a mono-exponential function, yielding a first order rate constant of $k_{red}$ = 130 ± 2 $s^{-1}$.

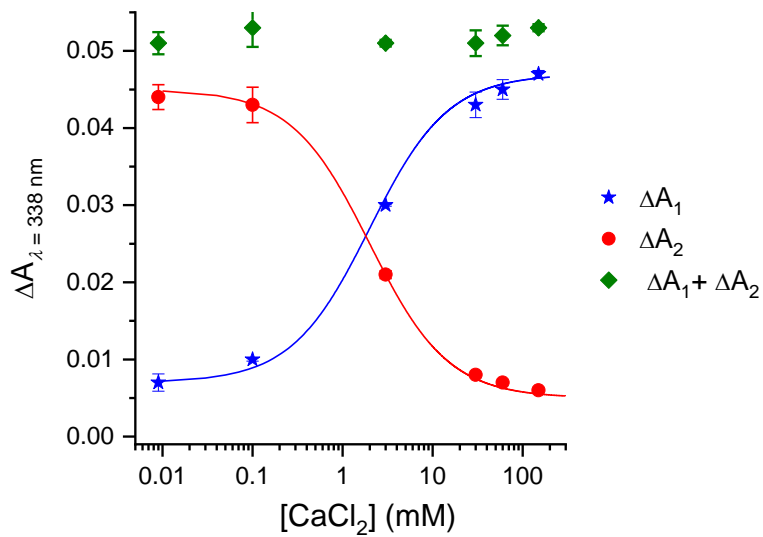

**Figure S3.** Effect of CaCl$_2$ concentration on the amplitude ΔA$_1$ and ΔA$_2$ associated to each of the two kinetic phases recorded during the transition of apo-sGDH to holo$_{red}$. The experimental conditions were the same than in Figure 1C. The amplitudes were inferred from the fit of a biexponential function to the stopped-flow kinetics traces. The blue and red solid lines correspond to the best fits of a binding isotherm to the blue and red experimental data, respectively, leading both to a same apparent equilibrium constant $K_d^{apo\text{-}sGDH/Ca} = 2.0 \pm 0.5$ mM.

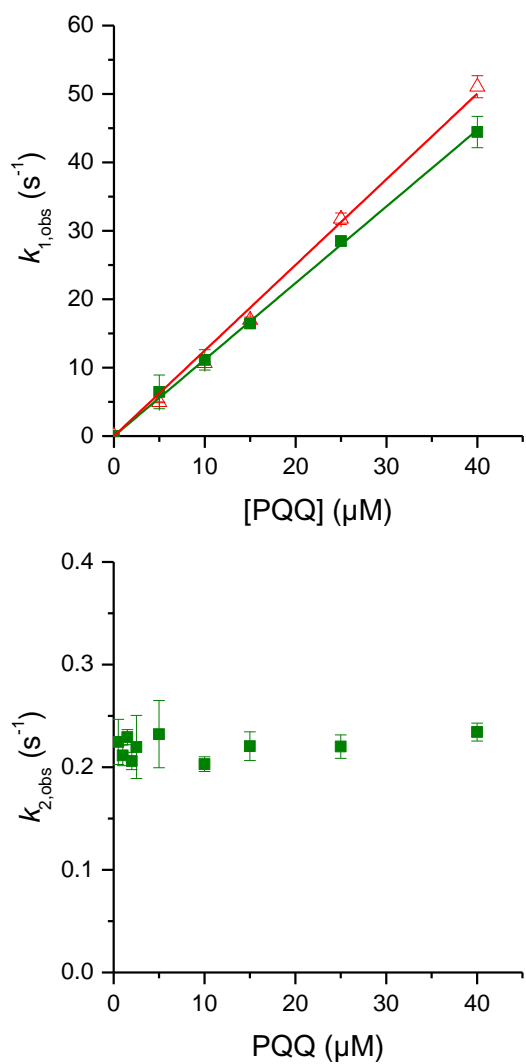

**Figure S4.** Plots of $k_{1,obs}$ and $k_{2,obs}$ as a function of PQQ concentration (for $k_{1,obs}$ only pseudo-first order conditions in PQQ were considered). The green square data were recovered from the absorbance change at 338 nm, while the red triangle data were obtained from the quenching of tryptophan emission fluorescence at 340 nm. The experimental conditions for absorbance change at 338 nm were the same than those reported in Figure 1A, while for the quenching of fluorescence they were the same than those in Figure 3A. Error bars are the standard deviations determined from the average of four consecutive stopped-flow kinetic experiments. Straight lines are the linear regression fits to the data. From the slopes of linear regression fits, the following bimolecular rate constants for the reaction between PQQ and apo-sGDH/$Ca^{2+}$ were determined: (green square) $k_{1,PQQ} = (1.12 \pm 0.01) \times 10^6$ $M^{-1}$ $s^{-1}$, (red triangle) $k_{1,PQQ} = (1.25 \pm 0.03) \times 10^6$ $M^{-1}$ $s^{-1}$.

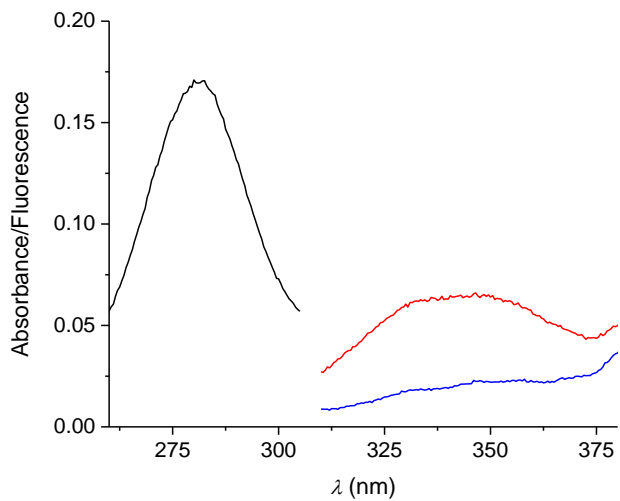

**Figure S5.** Fluorescence spectra of 5 μM apo-sGDH subunit before (red curve) and after addition of 10 μM PQQ (blue curve). The UV-visible absorption spectra of apo-sGDH is also added for comparison (black curve). The excitation was performed at $\lambda_{ex}$ = 297 nm.

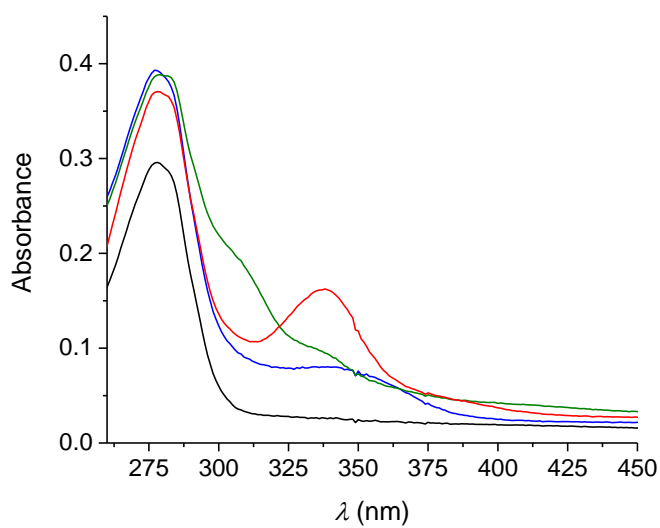

**Figure S6.** Absorption spectra (in 50 mM TRIS of pH 7.5 + 200 μM EDTA to scavenge any residual amount of $Ca^{2+}$ present in solution) obtained upon sequential conversion at room temperature of (black curve) apo-sGDH (5 μM subunit) to (blue curve) holoX$_{ox}$, followed by (green curve) holoX$_{red}$ converted then (red curve) to holo$_{red}$. After recording the spectrum of apo-sGDH, 2-fold excess of PQQ was added to the apo-sGDH solution leading instantaneously to the spectrum of holoX$_{ox}$ (blue curve). Then, 2 mM D-glucose was added to the holoX$_{ox}$ solution leading to a change in the spectrum (completed within a few minutes) corresponding to holoX$_{red}$ (green curve). Finally, 2 mM CaCl$_2$ was added to the holoX$_{red}$ solution leading rapidly to the spectrum of holo$_{red}$ (red curve).

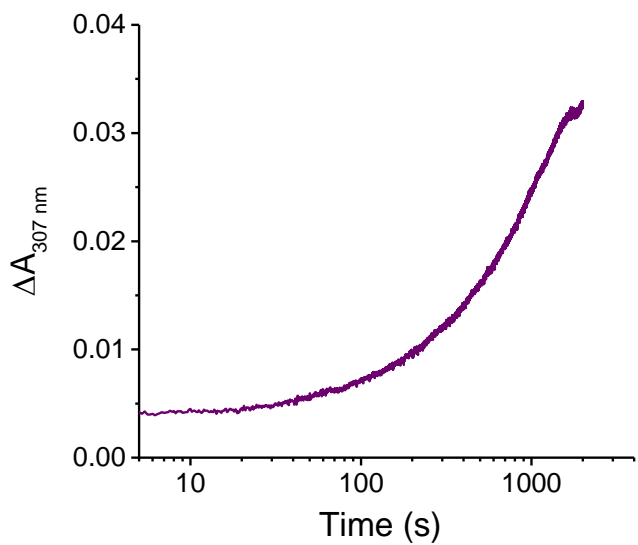

**Figure S7.** Kinetic trace for the reductive conversion of holoX$_{ox}$ to holoX$_{red}$ by glucose. The experiment was performed by mixing in a stopped flow apparatus 3.0 µM holoX$_{ox}$ subunit (prepared from a premix of 3.0 µM apo-sGDH subunit, 100 µM EDTA and 6 µM PQQ in a 50 mM TRIS buffer of pH 7.5) to a 200 µM glucose solution (prepared in the same 50 mM TRIS buffer of pH 7.5). The experiment was performed at 10°C and final concentrations were divided by 2 after mixing. From the fit of the data to a single exponential function a rate constant of $k_{red,X} = (8.4 \times 10^{-4} \pm 0.01) \times 10^{-4}$ s$^{-1}$ is inferred.

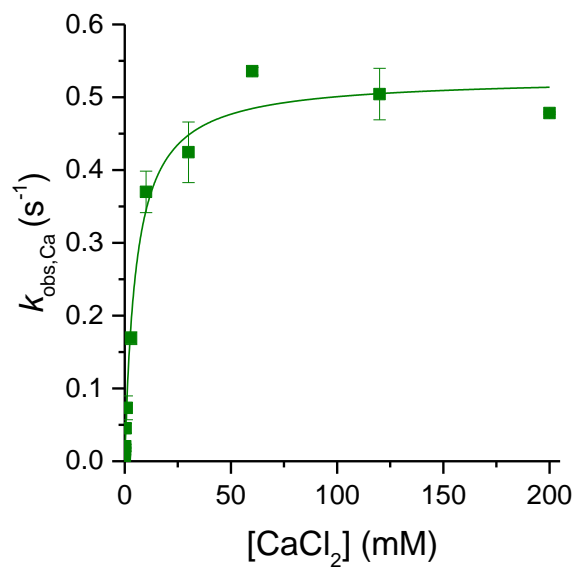

**Figure S8.** Influence of CaCl$_2$ on the observed rate related to the conversion of holoX$_{red}$ to holo$_{red}$. The experiments were performed at 10°C by mixing in a stopped flow instrument 5 µM holoX$_{red}$ subunit (prepared from 5 µM apo-sGDH, 10 µM PQQ, 200 µM EDTA and 2.5 mM glucose in a 50 mM TRIS of pH 7.5) and variable concentrations of CaCl$_2$ (ranging from 0.1 to 400 mM in 50 mM TRIS pH 7.5). After rapid mixing (final concentrations divided by 2), the time-course of holo$_{red}$ formation was recorded at 338 nm. For each CaCl$_2$ concentration, the observed rate $k_{obs,Ca}$ was recovered from the fit of a mono-exponential function to the experimental kinetic traces. The green solid line on the graph corresponds to the nonlinear regression of the following Michaelis-Menten-like kinetic equation $k_{obs,Ca} = \left(k'_{3,Ca}\left[Ca^{2+}\right]\right) / \left(K^M_{3,Ca} + \left[Ca^{2+}\right]\right)$ (where $k'_{3,Ca}$ is the maximal rate constant and $K^M_{3,Ca}$ is an apparent Michaelis-Menten constant) to the experimental data. From the best curve fitting the following values of $k'_{3,Ca} = 0.52 \pm 0.01$ s$^{-1}$ and $K^M_{3,Ca} = 5.3 \pm 0.7$ mM are obtained.

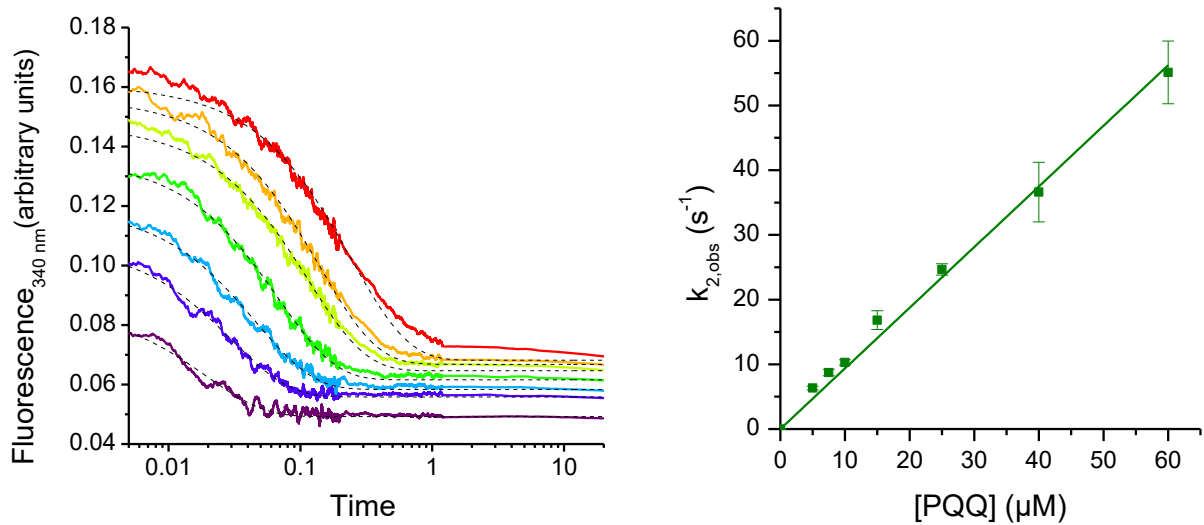

**Figure S9.** (Left) Time-course of the quenching of tryptophan emission fluorescence during the conversion of apo-sGDH to holoX (*i.e.,* in the absence of $Ca^{2+}$ and glucose) as a function of PQQ concentration. Experiments were performed by mixing in a stopped flow apparatus 5 µM apo-sGDH subunit in syringe 1 with different concentrations of PQQ in syringe 2. All solutions were prepared in a 50 mM TRIS buffer (pH 7.5) containing 200 µM EDTA in order to scavenge any residual amount of free $Ca^{2+}$ in solution. All experiments were also performed at 10°C. After rapid mixing (final concentrations divided by 2), the quenching of tryptophan emission fluorescence was followed at 340 nm with an excitation at 297 nm. (Right) Observed rate as a function of [PQQ]. The straight line is the linear regression fit to the data, from the slope of which the second order rate constant characterizing the reaction between PQQ and apo-sGDH to lead holoX is determined: $k_{2,PQQ} = (0.94 \pm 0.20) \times 10^6$ $M^{-1}$ $s^{-1}$.

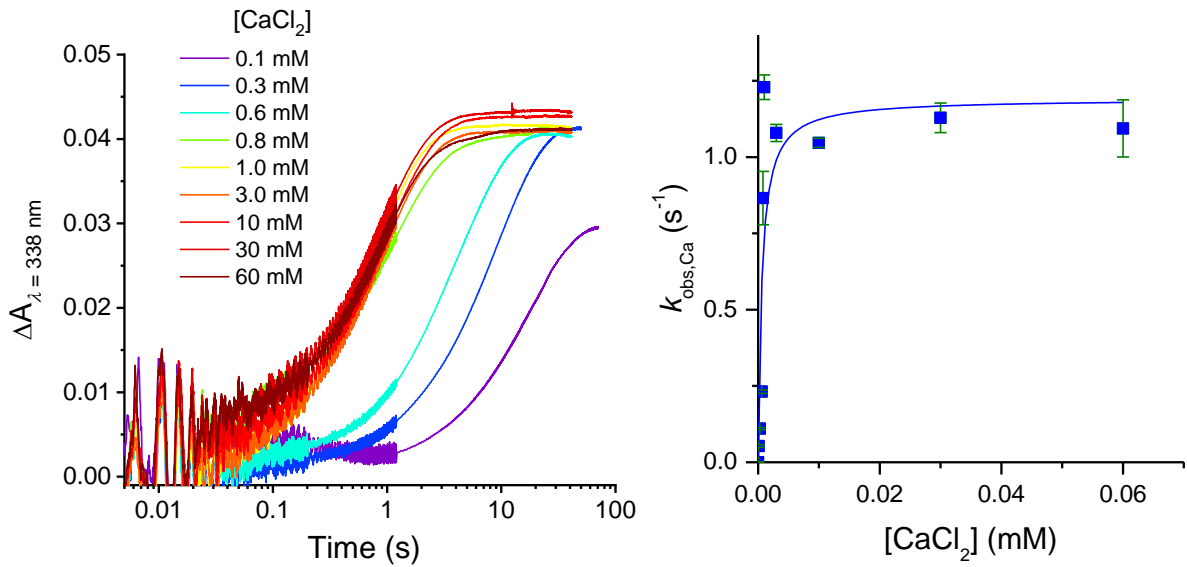

**Figure S10.** Influence of CaCl$_2$ on the observed rate related to the conversion of holoX$_{ox}$ to holo$_{red}$. The experiments were performed at 10°C by mixing in a stopped flow instrument 5 µM holoX$_{red}$ subunit (prepared from 5 µM apo-sGDH, 10 µM PQQ, 200 µM EDTA and 2.5 mM glucose in a 50 mM TRIS of pH 7.5) and variable concentrations of CaCl$_2$ (ranging from 0.1 to 400 mM in 50 mM TRIS pH 7.5). After rapid mixing (final concentrations divided by 2), the time-course of holo$_{red}$ formation was recorded at 338 nm. For each CaCl$_2$ concentration, the observed rate $k_{obs,Ca}$ was recovered from the fit of a mono-exponential function to the experimental kinetic traces. The green solid line on the graph corresponds to the nonlinear regression of the following Michaelis-Menten-type kinetic equation $k_{obs,Ca} = \left(k'_{2,Ca}\left[Ca^{2+}\right]\right) / \left(K^M_{2,Ca} + \left[Ca^{2+}\right]\right)$ (where $k'_{2,Ca}$ is the maximal rate and $K^M_{2,Ca}$ is an apparent Michaelis-Menten constant) to the experimental data. From the best curve fitting the following values of $k'_{2,Ca} = 1.2 \pm 0.2$ s$^{-1}$ and $K^M_{2,Ca} = 0.6 \pm 0.3$ mM are obtained.

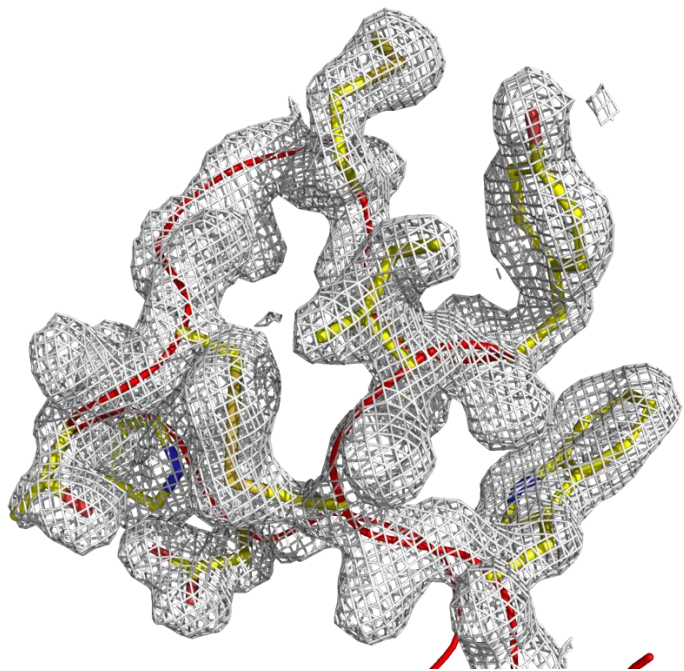

**Figure S11.** Experimental electron density map (2Fo-Fc, 1.2σ) superimposed to the apo-sGDH (5MIN) Cα trace in the region of the loop 334-346. Key residues are shown in stick mode.

**Table S1.** Extinction coefficients (at different selected wavelengths) of the species involved in the reconstitution of sGDH.

| Species | $\lambda$ (nm) | $\varepsilon$ (M$^{-1}$ cm$^{-1}$) |
| --- | --- | --- |
| holo$_{ox}$ | 277 | 85000 |
| | 338 | 12900 |
| holo$_{red}$ | 277 | 76800 |
| | 338 | 32400 |
| holoX$_{ox}$ (or I$_{ox}$)[a] | 338 | 16000 |
| | 307 | 19000 |
| holoX$_{red}$ (or I)[a] | 338 | 19000 |
| | 307 | 39000 |
| PQQ free (or PQQ/Ca)[a] | 338 | 12300 |
| apo-sGDH | 277 | 60000 |
| | 338 | 5000 |

[a] It was assumed that the extinction coefficient of these species is the same either there is or not the presence of Ca$^{2+}$

**Table S2.** Crystallographic data collection and model refinement statistics for the apo-sGDH enzyme.

| Data Collection | |
|---|---|
| PDB ID | 5MIN |
| Wavelength [Å] | 1.54187 |
| Temperature [K] | 100 |
| Space group | P(1) |
| Unit Cell Parameters | |
| a≠b≠c [Å] | 54.407 54.898 85.474 |
| α≠ß≠γ [°] | 88.474 81.58 69.62 |
| Resolution range [Å] [a] | 25.73 – 1.76 |
| Reflections [a] | |
| Completeness [%] | 92.5 (70.8) |
| Data quality [a] | |
| Intensity [I/σ(I)] | 9.49 (2.47) |
| R meas [%] [b] | 5.55 |
| Wilson B value [Å$^2$] | 14.8 |
| **Refinement** | |
| Resolution range [Å] [a] | 25.73 – 1.76 |
| Reflections [a] | |
| Number | 84103 |
| Test Set (5%) | 4137 |
| R work[a,c] | 0.169 |
| R free [a,d] | 0.204 |
| Contents of the Asymmetric Unit | |
| Protein: Molecules, Residues, Non-Hydrogen Atoms | 8241 |
| Ligands: (Molecules) | 7 |
| Water: Molecules | 1099 |
| RMSD [e] | |
| Bond Lengths [Å] | 0.5 |
| Bond Angles [°] | 0.64 |
| Validation Statistics | |
| Ramachandran Plot | |
| Residues in Allowed Regions [%, No.] | 3 |
| Residues in Favoured Regions [%, No.] | 97 |
| Outliers [residues] | 4 |

[a] Data for the highest resolution shell in parenthesis

[b] R meas = $\sum h [ n / n − 1 ] 1/2 \sum i |I h − I h,i | / \sum h \sum i I h,i$ , where I h is the mean intensity of symmetry-equivalent reflections and n is the redundancy

[c] R work = $\sum h | F o − F c | / \sum F o$ (working set, no σ cut-off applied)

[d] R free is the same as R work , but calculated on 5 % of the data excluded from refinement

[e] Root-mean-square deviation from target geometries

**Table S3.** Observed rates and absorbance difference amplitudes extracted from the fits of a bi-exponential function to the experimental kinetics plots of Figure 1A.

| [PQQ] (µM) | $k_{1,obs}$ (s$^{-1}$) | $\Delta A_1$ | $k_{2,obs}$ (s$^{-1}$) | $\Delta A_2$ | $\Delta A_{max}$ | $\Delta A_1/\Delta A_{max}$ |
|---|---|---|---|---|---|---|
| 0.5 | 2.0 | 0.007 | 0.16 | 0.003 | 0.010 | 0.70 |
| 1 | 2.5 | 0.012 | 0.18 | 0.006 | 0.018 | 0.67 |
| 1.5 | 2.7 | 0.016 | 0.18 | 0.010 | 0.026 | 0.62 |
| 2 | 2.9 | 0.021 | 0.21 | 0.015 | 0.036 | 0.58 |
| 2.5 | 3.1 | 0.024 | 0.19 | 0.021 | 0.045 | 0.53 |
| 5 | 4.1 | 0.030 | 0.23 | 0.021 | 0.051 | 0.59 |
| 10 | 9.6 | 0.032 | 0.23 | 0.019 | 0.051 | 0.62 |
| 15 | 15.9 | 0.030 | 0.24 | 0.020 | 0.050 | 0.60 |
| 20 | 25.7 | 0.030 | 0.28 | 0.020 | 0.050 | 0.60 |
| 40 | 41.8 | 0.030 | 0.29 | 0.023 | 0.053 | 0.57 |

**Table S4.** Observed rates and absorbance difference amplitudes extracted from the fits of a bi-exponential function to the experimental kinetics plots of Figure 1C.

| [Ca$^{2+}$] (mM) | $k_{1,obs}$ (s$^{-1}$) | $\Delta A_1$ | $k_{2,obs}$ (s$^{-1}$) | $\Delta A_2$ | $\Delta A_{max}$ | $\Delta A_1/\Delta A_{max}$ |
|---|---|---|---|---|---|---|
| 0.009 | 1.26 | 0.007 | 0.01 | 0.044 | 0.051 | 0.14 |
| 0.1 | 2.75 | 0.010 | 0.04 | 0.043 | 0.053 | 0.19 |
| 3 | 4.11 | 0.030 | 0.23 | 0.021 | 0.051 | 0.58 |
| 30 | 2.18 | 0.043 | 0.44 | 0.008 | 0.051 | 0.84 |
| 60 | 1.11 | 0.045 | 0.24 | 0.007 | 0.052 | 0.86 |
| 150 | 0.48 | 0.047 | 0.11 | 0.006 | 0.053 | 0.89 |

**Table S5.** Observed rate and fluorescence difference amplitude extracted from the fits of a mono-exponential function to the experimental kinetics plots of Figure 3B.

| [Ca$^{2+}$] (mM) | $k_{obs}$ (s$^{-1}$) | $\Delta F$ [a] |
|---|---|---|
| 0.009 | 4.14 | 0.109 |
| 0.1 | 5.22 | 0.110 |
| 3 | 4.78 | 0.111 |
| 30 | 1.98 | 0.105 |
| 60 | 1.10 | 0.099 |
| 150 | 0.45 | 0.093 |

[a] fluorescence difference amplitude

**Table S6.** Comparison of the observed rate constants $k_{1,obs}$ and $k_{2,obs}$ determined for the wild type and P248A mutant sGDH under different stopped-flow experimental conditions.

| Stopped-flow Experiment | | 6 mM $CaCl_2$ + 5 μM apo-sGDH | 5 μM apo-sGDH | 3 mM $CaCl_2$ + 5 μM apo-sGDH |
|---|---|---|---|---|
| | **Syringe 1** | | | |
| | **Syringe 2** | 10 μM PQQ | 6 mM $CaCl_2$ + 10 μM PQQ | 3 mM $CaCl_2$ + 10 μM PQQ |
| **Type of apo-sGDH** | | Recovered rate constants | | |
| Wild-type (UV-vis absorbance at 338 nm) | $k_{1,obs}$ (s$^{-1}$) | 4.94 ± 0.14 | 4.78 ± 0.22 | 6.45 ± 2.46 |
| | $k_{2,obs}$ (s$^{-1}$) | 0.25 ± 0.03 | 0.22 ± 0.01 | 0.23 ± 0.03 |
| Wild-type (quenching of fluorescence) | $k'_{1,obs}$ (s$^{-1}$) | 5.68 ± 0.25 | 6.51 ± 0.30 | 4.94 ± 0.06 |
| P248A (UV-vis absorbance at 338 nm) | $k_{1,obs}$ (s$^{-1}$) | 6.60 ± 1.21 | 17.13 ± 3.11 | 13.22 ± 1.88 |
| | $k_{2,obs}$ (s$^{-1}$) | 2.00 ± 0.16 | 5.37 ± 1.94 | 3.70 ± 0.41 |

In these experiments, the pre-incubation of either apo-sGDH (wild-type or mutated sGDH) with $CaCl_2$ in syringe 1 or PQQ with $CaCl_2$ in syringe 2, or even in both syringes before mixing, were tested to verify if there is an effect on the observed rates $k_{1,obs}$ and $k_{2,obs}$ determined for the wild-type and the P248A mutant-type sGDH. This was achieved by monitoring either the enzyme activation through the absorbance change at 338 nm or the tryptophan fluorescence quenching at 335 nm. The final concentrations for all experiments were 2.5 μM apo-GDH subunit, 5 μM PQQ and 3 mM $CaCl_2$ in 50 mM TRIS buffer pH 7.5. The temperature was maintained at 10°C.

Graphical Abstract

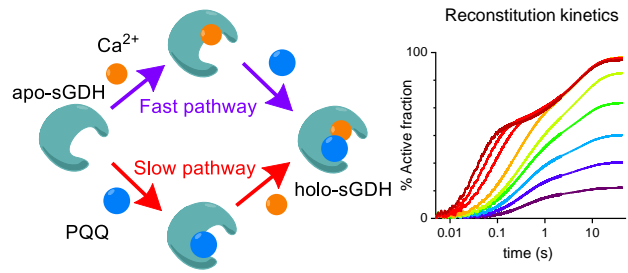